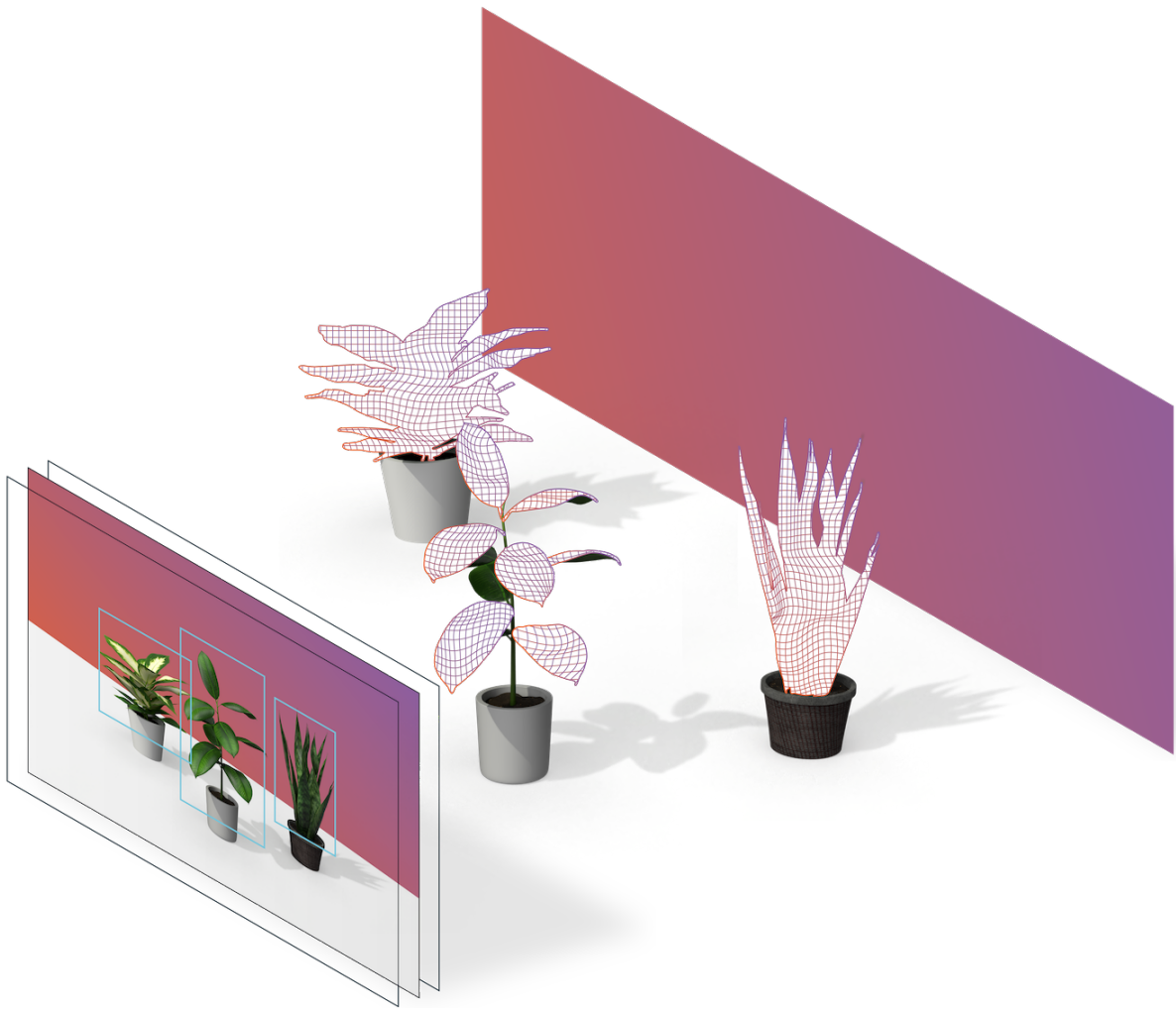

# Do You See What I See?

**Capabilities and Limits of Automated Multimedia Content Analysis**

**Carey Shenkman**
**Dhanaraj Thakur**
**Emma Llansó**

**May 2021**

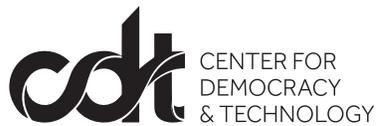

The Center for Democracy & Technology (CDT) is a 25-year-old 501(c)3 nonpartisan nonprofit organization working to promote democratic values by shaping technology policy and architecture. The organisation is headquartered in Washington, D.C. and has a Europe Office in Brussels, Belgium.---

## CAREY SHENKMAN

Carey Shenkman is an independent consultant and human rights attorney.

## DHANARAJ THAKUR

Dhanaraj Thakur is the Research Director at CDT, where he leads research that advances human rights and civil liberties online.

## EMMA LLANSÓ

Emma Llansó is the Director of CDT's Free Expression Project, where she leads CDT's work to promote laws and policies that support Internet users' free expression rights in the United States, Europe, and around the world.

# Do You See What I See?

## Capabilities and Limits of Automated Multimedia Content Analysis

**Carey Shenkman**
**Dhanaraj Thakur**
**Emma Llansó**

**WITH CONTRIBUTIONS BY**

DeVan Hankerson, Hannah Quay-de la Vallee, Samir Jain, and Tim Hoagland.

**ACKNOWLEDGEMENTS**

We thank Robin Burke for his feedback on sections of this paper. We also thank the various experts from academia, industry, and civil society that we interviewed and who helped inform the analysis in this paper.

This work is made possible through a grant from the John S. and James L. Knight Foundation.

**Suggested Citation:** Shenkman, C., Thakur, D., Llansó, E. (2021) Do You See What I See? Capabilities and Limits of Automated Multimedia Content Analysis. Center for Democracy & Technology.

# Executive Summary

The ever-increasing amount of user-generated content online has led, in recent years, to an expansion in research and investment in automated content analysis tools. Scrutiny of automated content analysis has accelerated during the COVID-19 pandemic, as social networking services have placed a greater reliance on these tools due to concerns about health risks to their moderation staff from in-person work. At the same time, there are important policy debates around the world about how to improve content moderation while protecting free expression and privacy. In order to advance these debates, we need to understand the potential role of automated content analysis tools.

This paper explains the capabilities and limitations of tools for analyzing online multimedia content and highlights the potential risks of using these tools at scale without accounting for their limitations. It focuses on two main categories of tools: matching models and computer prediction models. Matching models include cryptographic and perceptual hashing, which compare user-generated content with existing and known content. Predictive models (including computer vision and computer audition) are machine learning techniques that aim to identify characteristics of new or previously unknown content.

These tools are most useful under certain conditions:

- **Matching models** are generally well-suited for analyzing known, existing images, audio, and video, particularly where the same content tends to be circulated repeatedly.
  - Perceptual hashing is almost always better-suited to matching items that feature slight variations, which may occur either naturally or from attempts to circumvent detection.
- **Predictive models** can be well-suited to analyzing content for which ample and comprehensive training data is available. They may also perform well in identifying objective features in multimedia. Examples may include whether multimedia contains clear nudity, blood, or discrete objects.
  - Analysis of static images is much more straightforward than video analysis.
  - Analysis of audio often involves a two-step process of transcription followed by analysis of the transcribed text.

Even in these scenarios, automated multimedia content analysis tools have many limitations. And those limitations become even more evident when the tools are used in more challenging settings. Any applications of these tools should consider at least five potential limitations:

# 1. Robustness

**State-of-the-art automated analysis tools that perform well in controlled settings struggle to analyze new, previously unseen types of multimedia.**

Automated models are repeatedly shown to fail in situations they have never encountered in their design or training. *Robustness* of the tools underlying automated content analysis—or the ability to not be fooled by minor distortions in data—is a constant and unsolved problem. Some challenges for automated analysis are due to natural occurrences (such as a photograph taken at a slightly different angle from a reference photo). But in a social media analysis setting, many challenges are *deliberately* caused by efforts to slip past detection. These can include anything from watermarks, to subtle rotations or blurs, to sophisticated methods such as deepfakes which create synthetic, realistic-seeming videos to harass or spread disinformation. Machine learning models struggle with these cases because circumvention efforts are constantly evolving, and models may be over-optimized for the examples with which they are created or trained. They may not generalize performance well to novel data. This is akin to memorizing answers to specific questions before a test without actually understanding the underlying concepts.

# 2. Data Quality

**Decisions based on automated content analysis risk amplifying biases present in the real world.**

Machine learning algorithms rely on enormous amounts of training data, which can include large databases of photos, audio, and videos. It is well documented that datasets are susceptible to both intended and unintended biases. How specific concepts are represented in images, videos, and audio may be prone to biases on the basis of race, gender, culture, ability, and more. Multimedia sampled randomly from real-world data can likewise propagate real-world biases. For example, existing news coverage of "terrorist propaganda" often perpetuates racial and religious biases. This can lead to problematic asymmetries as to what automated models identify as "terrorist" images. While some methods exist for attempting to mitigate these biases at the machine learning level, they are far from sufficient. Moreover, efforts to "clean" datasets to address some kinds of risks can actually introduce other forms of bias into the training data.

## 3. Lack of Context

**Automated tools perform poorly when tasked with decisions requiring appreciation of context.**

While some types of content analysis may be relatively straightforward, the task of understanding user-generated content is typically rife with ambiguity and subjective judgment calls. Certain types of content are easier to classify without *context*—i.e. there may be wider consensus on what constitutes gore, violence, and nudity versus what is sexually suggestive or hateful. And even then, for instance, artistic representations and commentary may contain nudity or violence but be permitted on a given service when depicted in those contexts. The same content shared by one person in a particular setting, such as photos of baked goods, may have entirely different implications in another where those baked goods are a photo selling illicit drugs. Machines are ill-suited to make contextual assessments or apply the nuanced ethical standards that may be necessary for any given decision.

## 4. Measurability

**Generalized claims of accuracy typically do not represent the actual multitude of metrics for model performance.**

Real-world impacts of automated analysis decisions may be difficult or impossible to measure without knowing all the content a system fails to properly analyze. For this and other reasons, metrics that convey reliability in the content analysis space, such as "99.9% accuracy," are typically practically meaningless. For example, some forms of harmful content, such as terrorist propaganda, can comprise a very small percentage of multimedia content. An algorithm that merely labels every piece of content "not extreme" could technically be "accurate" at least 99.9% of the time. But it would be right *for entirely the wrong reasons*. Moreover, even if a model predicted the right result 999 out of 1000 times, the one wrong result might have extremely harmful impacts at a scale of millions or billions of pieces of content. Metrics of positive model performance may also be self-selective. They may result from optimization to a specific dataset that is not generalizable to real-world problems. Better measures than "accuracy" are metrics such as *precision* and *recall*, which capture false negative and false positive rates.

## 5. Explainability

**It is difficult to understand the steps automated tools take in reaching conclusions, although there is no "one-size-fits-all" approach to explainability.**

State-of-the-art machine learning tools, by default, cannot be "opened up" to get a plain-spoken explanation of why they reached a decision they did. These tools utilize large *neural networks* which may have up to millions or billions of interrelated parameters involved in learning and producing outputs. While the inputs and outputs of these systems may be understood by humans, comprehending the intermediate steps, including how an automated analysis system makes decisions or weighs various features, is a daunting technical task, and these intermediate steps typically do not translate into the kinds of judgments a human would make. Research efforts are being made to promote *explainability*, the ability to map the operations of machine judgment onto concepts that can be understood by humans. Explainability has important implications for developing trust in these systems and for preventing disparate impacts across various groups, as well as identifying opportunities for redress. At the same time, explainability may vary depending on whether what needs to be known involves the factors in a singular decision, or the structural characteristics of a network as a whole.

While there are many important and useful advances being made in the capabilities of machine learning techniques to analyze content, policymakers, technology companies, journalists, advocates, and other stakeholders need to understand the limitations of these tools. A failure to account for these limitations in the design and implementation of these techniques will lead to detrimental impacts on the rights of people affected by automated analysis and decision making. For example, a tool with limited robustness can be circumvented and fail to identify abusive content. Poor data quality can lead to machine learning models that perpetuate or even exacerbate existing biases in society, and can yield outputs with a disparate impact across different demographics. Insufficient understanding of context can lead to overbroad limits on speech and inaccurate labeling of speakers as violent, criminal, and abusive. Poor measures of the accuracy of automated techniques can lead to a flawed understanding of their effectiveness and use, which can lead to an over-reliance on automation and inhibit the introduction of necessary safeguards. Finally, limited explainability can restrict the options for remedying both individual errors and systematic issues, which is particularly important where these tools are part of key decision-making systems.

Large scale use of the types of automated content analysis tools described in this paper will only amplify their limitations and associated risks. As a result, such tools should seldom be used in isolation; if they are used, it should only be as part of more comprehensive systems that incorporate human review and other opportunities for intervention. Design of such systems requires an accurate understanding of the underlying tools being used and their limitations.

Policymakers must also be versed in the limitations of automated analysis tools to avoid promulgating statutes or regulations based on incorrect assumptions about their capabilities. For example, legislators should not pass laws about content moderation that are premised on the ability of automated analysis tools to perform moderation tasks at scale, and automated content filtering should never be required by law. More generally, policies that do not account for the limitations discussed here risk normalizing an uncritical view of the efficacy of these tools. This can undermine important and needed public dialogue about what problems machine learning or "artificial intelligence" can – and cannot – help us solve.

# Introduction

The sheer scale of uploaded user-generated content (UGC) has increased dramatically in recent years, leading to an explosion in the research about and use of automated techniques to analyze and moderate it (Cambridge Consultants, 2019). The COVID-19 pandemic triggered a "massive experiment" in algorithmic content moderation, driven in large part by social distancing requirements that meant the human workforces in charge of content analysis were sent home (Faddoul, 2020; Llansó, 2020b; Matsakis & Martineau, 2020). "It did not go well," reported *Politico* months later, observing the shortcomings of automated tools that simultaneously led to dramatically higher numbers of takedowns, while properly detecting far less questionable content. "Nobody appreciated the content moderators until they were gone"(Scott & Kayali, 2020).

The Center for Democracy and Technology (CDT) has closely followed the use of automated content analysis tools, both analyzing their potential value and their implications for human rights and freedom of expression. In 2017, CDT explained the limitations of natural language processing (NLP) tools for analyzing the *text* of social media posts and other online content in order to better help civil society, industry, and policymakers understand the available tools, as well as the strengths and weaknesses of using them (Duarte et al., 2017). In this study, we provide an accessible technical explanation of tools for analyzing *multimedia* content — images, audio, and video, as well as live streamed video. This study hence focuses on a subset of analysis tools that present unique technical challenges.

Policymakers worldwide are increasingly calling on social media companies to identify and restrict text, photos, and videos that involve illegal, harmful, or false information (Browning, 2020; Wong, 2020). Many services are voluntarily incorporating automation into their content moderation systems, and government agencies are also exploring the use of automated content analysis. Countries around the world are also proposing legal mandates for companies to filter content or to respond to takedown orders within very short time frames, which apply significant pressure on these companies to employ automation. Understanding these tools, and their capabilities and limitations when used in connection with multimedia, is crucial for stakeholders to make informed choices: users engaged with and affected by social media; companies weighing appropriate technologies and safeguards; policymakers determining whether to enact laws and regulations that require, prohibit, or regulate the use of automated analysis tools; and civil society and journalists seeking to understand the implications of automated tools for content analysis.

The first part of this paper discusses tools that are used for automated analysis of multimedia content. The second part of the paper discusses five limitations of these tools that policymakers and developers should understand when considering the role these tools may play in the analysis and moderation of user-generated content:

1. **Robustness.** State-of-the-art automated analysis tools that perform well in controlled settings struggle to analyze new, previously unseen types of multimedia.

2. **Data Quality.** Decisions based on automated multimedia content analysis risk amplifying biases present in the real world.

3. **Lack of Context.** Automated tools perform poorly when tasked with decisions requiring judgment or appreciation of context.

4. **Measurability.** Generalized claims of accuracy typically do not represent the actual multitude of metrics for model performance.

5. **Explainability.** It is difficult to understand the steps automated tools take in reaching conclusions, although there is no "one-size-fits-all" approach to explainability.

This paper concludes with a discussion of the implications and risks of these limitations for relevant stakeholders, including civil society, industry, and policymakers.

**Countries around the world are also proposing legal mandates for companies to filter content or to respond to takedown orders within very short time frames, which apply significant pressure on these companies to employ automation. Understanding these tools, and their capabilities and limitations when used in connection with multimedia, is crucial for stakeholders to make informed choices.**

# I. Tools for Automated Multimedia Content Analysis

Content analysis requires perception, recognition, and judgment. But human visual and auditory perception has had hundreds of thousands of years to evolve, and involves judgments that are the result of years of education and socializing. Imagine that you were asked to identify whether a picture contained a dog, but you had never seen a dog, or any animal, before in your life. How would you go about this task? Perhaps you are shown photos of dogs and learn to recognize them. You could recognize if you were shown the same photo twice. But this does not help you evaluate new photos. You could learn to associate that an animal that stands on four legs and has a tail is a dog. But this rule will not help you get the right answer if you are shown a photo of a horse. You may end up developing a very clear understanding of what a "dog" is, but unless you had additional specific training, you may not be able to differentiate labradors from golden retrievers. We can take for granted the years of learning that have enabled us to make these determinations. These are just some of the challenges that we are asking computers to address through machine learning.

**Machine learning (ML)** is a process by which a system parses data to extract characteristics, relationships, and correlations from it, without being programmed to do so, and then applies those understandings to analyze other data. The notion of machine learning dates back to 1952, but modern processing power has exponentially increased its potential. "Machine learning is a thing-labeler, essentially" (Kozyrkov, 2018). A subfield of machine learning called **deep learning** has accelerated and been the center of focus in the last several years for methods in vision and audition.

A machine can generally make identifications, or label things, in one of two ways: by **matching**, or recognizing something as identical or sufficiently similar to something it has seen before; or by **prediction**, recognizing the nature of something based on the machine's prior learning. The latter category gives rise to the fields of **computer vision** and **computer audition**, which respectively study how computers might achieve high-level understanding from images or videos, and audio. In the following sections, this paper will explore these concepts in depth.

Gorwa et al. (2020) point out that some technologies combine matching and predictive techniques, such as facial recognition technology that identifies matches of previously identified faces and also attempts to learn characteristics of faces in general.

 

# A. Matching Models for Multimedia Content Analysis

The simplest approach to content analysis is matching. A matching algorithm seeks to answer: "Have I seen this image, audio, or video before?" It enables an operator to compare a piece of UGC to a pre-existing library of content. Matching could hypothetically take place by comparing every pixel or bit of an image or video to another in order to find a perfect match, or listening to every fraction of a second of audio. But this would be computationally intensive, unfeasible, and easy to circumvent. Instead, content can be reduced to simpler representations to make comparison more efficient and flexible.

One way to do this is by **hashing**, which creates digital fingerprints of content in order to produce a significantly more compact and manageable object for comparison, while maintaining semantic accuracy. Similar to the way a person can be identified using a physical fingerprint, so too can pieces of digital content be identified by their corresponding digital fingerprints (Singh, 2019). Hash functions can be cryptographic or perceptual.

**Cryptographic hashing** uses a cryptographic function to generate a random hash fingerprint, which is extremely sensitive to change. For example, changing the shade of one pixel in a high-resolution photo would produce a distinct cryptographic hash. This can be highly effective in authenticating known content without alterations.

These same cryptographic functions are what encode encrypted messages. There, they also serve as a guarantee that not a single bit (or letter or word) in the message has been changed.

**Perceptual hashing**, on the other hand, seeks to determine not whether two pieces of content are identical, but whether they are "alike enough"—i.e. practically identical. Perceptual hashing methods utilize algorithms to better comprehend the nature of a piece of content so that minor changes cannot fool the system. The operator of a system that uses perceptual hashes can set a threshold to determine what degree of difference between hashes is allowed to still consider them matches. Some specific implementations of perceptual hash algorithms include the detection of child sexual abuse material (CSAM), terrorist propaganda, and copyrighted content (see the Appendix for a more detailed description of these techniques and how they are applied in content analysis).

One important property of hashing is that it requires knowing in advance the content to be identified. To use human fingerprints to identify humans, one needs a database of the fingerprints of known individuals to reference against. Similarly, to use image hashes to detect unwanted content, operators must have a database of reference content to be matched against.

There are at least two ways to assess the effectiveness of an algorithmic technique like a hashing algorithm:

- How **robust** is the function to natural or adversarial interference? A method may be able to better resist some forms of manipulations and distortions (including geometric, chromatic, noise, or blur) than others.

- How **discriminative** is the function? Discrimination represents the ability to correctly differentiate between distinct pieces of content. To be usable, an algorithm needs to avoid reporting matches of distinct content—i.e. avoid false positives (Martínez et al., 2018).

Highly robust models have a low rate of false negatives, and highly discriminative models have a low rate of false positives. Related concepts are a model's positive predictive value, known as **precision**, and true positive rate, known as **recall** (Drmic et al., 2017). Precision is the ratio of true positive results to all positives predicted by the model (including false positives). It is an important measure to use in instances where the cost of a false positive is high. Recall refers to the ratio of true positives to all actual positives in the sample (including false negatives). The recall of a model is important in instances where the cost of a false negative is high. Thus, depending on context measures for either, recall or precision may be more relevant.

## STRENGTHS, WEAKNESSES, AND CONSIDERATIONS OF MATCHING MODELS

The most significant characteristic of matching for multimedia content analysis is that it will only consider matches to content that is already contained in a reference database. Thus, it cannot be used for new content—beyond minor manipulations of existing content—that has not been provided for in the database. One problem with this approach is where content in the reference database is objectionable or even illicit, which means that maintaining those references may raise ethical or legal concerns. Matching technologies are most effective in categories of content that are predisposed to sharing already-known multimedia.

Depending on how matching algorithms are deployed, they may be designed in a way to prioritize the minimization of false positives (Du et al., 2020). Comparing hashes may also not require significant computational resources (Engstrom & Feamster, 2017). Further, notwithstanding concerns regarding transparency of hash databases, the decision-making process of matching algorithms is relatively straightforward and explainable.

PhotoDNA, developed by Microsoft, is presently the most widespread perceptual matching method for countering child sexual abuse material (CSAM). Some of its main advantages are its low false-positive detection rate, its relatively low computational costs, and resistance against reverse-engineering attacks to identify individuals in images (Nadeem et al., 2019; Pereira et al., 2020). Various other perceptual hashing methods have shown adaptability to recognizing multimedia in varied settings. For instance, Echoprint (an open-source fingerprinting library utilized by Echo Nest, a subsidiary of Spotify) is flexible enough to identify remixes, live versions, and sometimes covers of music (Brinkman et al., 2016). However, changes and "noise" of various forms can still present challenges for state-of-the-art algorithms. Google utilizes machine learning in its Now Playing audio fingerprint, which can be used to identify ambient music, though its effectiveness is dependent on the type of noise in the background environment (Agüera y Arcas et al., 2017; Lyon, 2018). Facebook's open-source PDQ and TMK+PDQF algorithms for image and video hashing, respectively, both perform strongly against certain types of changes like minor/imperceptible changes in content, but struggle with more deliberate or major changes like the addition of watermarks (Dalins et al., 2019).

An important consideration in utilizing matching-based systems is their general inability to assess context. The same pieces of content that are objectionable in one context may have significant expressive and public interest value in a different setting, such as in art, academic or journalistic work, or human rights commentary. In the area of copyright, "fair use" is a recognized allowance for the dissemination of copyrighted content. However, YouTube's Content ID, which allows rights holders to create fingerprints of their multimedia content, has generated controversy among legal scholars for the tension it creates with safe harbors in copyright law such as fair use. "The inability to recognize fair use is an issue inherent in automated filters like the Content ID system,"(Solomon, 2015, p. 21; see also Bartholomew (2014) and Trendacosta (2020)). These systems, known as an automated identification and remediation system (AIRS), are challenged by matching technology's failure to ascertain context (Zernay & Hagemann, 2017). Similarly, using hashes to identify and block content that may, in some contexts, be deemed "terrorist propaganda" can fail to account for situations where such content is being shared to document human rights abuses or to provide journalistic coverage (Human Rights Watch, 2020).

Facebook also maintains the membership-based ThreatExchange API, a clearinghouse for security-related information, and shares obtained hashes with GIFCT (Pham, 2019).

The way that hash databases are designed, maintained, and implemented can also amplify (or mitigate) the risks of hash-based analysis systems. For example, a hash database might be a shared resource to which multiple services have access, such as the database of child sexual abuse material maintained by the National Center for Missing and Exploited Children (NCMEC) or the database for terrorist propaganda content maintained by the Global Internet Forum to Counter Terrorism (GIFCT).

In the case of the GIFCT, each participating company may individually nominate content for inclusion in the database. Without clear parameters, the standards applied by each company, and thus to individual pieces of hashed content, may vary widely (Llansó, 2016). As CDT has highlighted previously, the practice of sharing definitions of prohibited content carries multiple risks, including promoting cross-platform censorship and imposing de facto standards of speech online (Llansó, 2020c; see also Douek (2020) and Radsch (2020)).

Hash databases may also present case-specific security risks. For instance, they can be susceptible to *hash collision* attacks or *hash poisoning*, wherein an attacker reverse engineers an image from a given hash, and deliberately introduces content to generate false positives (Dolhansky & Ferrer, 2020). If that content becomes included in the underlying database, then it would potentially serve to allow outsiders to blacklist content for a variety of malicious objectives. The susceptibility of a hash function to such attacks depends in part on its predictability. In practice, this type of attack may generally require some knowledge of or access to the hash function.

### KEY TAKEAWAYS REGARDING MATCHING MODELS

- Matching models are well-suited for analyzing *known*, *existing* images, audio, and video.

- There are two main types of hashing methods, *cryptographic* and *perceptual*. Of the two, perceptual hashing is almost always better-suited to content analysis applications, which generally involve matching items that feature slight variations that may occur either naturally or from attempts to circumvent detection.

- Two metrics for measuring perceptual hashing are *robustness* and *discrimination*. Robustness refers to the ability of an algorithm to ignore changes that are perceptually irrelevant, i.e. minor changes that do not impact what a human would ultimately see. Discrimination refers to the ability to distinguish images or other content that is actually different.

- Key existing use cases for matching-based analysis are instances where the content is known and tends to be circulated repeatedly. These include child exploitation materials, terrorist propaganda, and copyrighted multimedia.

- Matching models require the maintenance of a database of images, audio, or video to which content can be compared. Where the material reflected in a hash database is objectionable or even illicit, maintaining those reference files may raise ethical or legal concerns; without those reference files, however, it is impossible to verify the contents of the hash database.

**An important consideration in utilizing matching-based systems is their general inability to assess context. The same pieces of content that are objectionable in one context may have significant expressive and public interest value in a different setting, such as in art, academic or journalistic work, or human rights commentary.**

# B. Predictive Models for Multimedia Content Analysis

Unlike matching algorithms, which attempt to authenticate a piece of content by assessing its similarity to an *existing* and *known* piece of content, predictive models aim to be able to identify characteristics of a *new* and *previously unseen* piece of content. To do this, a model must be able to generalize the attributes of whatever it seeks to classify. Prediction is a problem tackled in the areas of computer vision and computer audition.

Computer vision refers to techniques used to address a range of tasks in content analysis including analyzing shapes, textures, colors, spatial arrangement, and static and temporal relationships. Examples of computer vision tools include:

See the Appendix to this report for a more detailed discussion of each of these techniques.

**Classifiers.** These are algorithms that predict what an image contains. Image classifiers are one of the simpler computer vision tools, and they are among the most common in the multimedia content analysis space (Batra, 2019). A very basic example of a classifier would be one that predicts whether or not an image contains a cat or dog. While they may perform well in many domains, they are susceptible to external forms of visual interference and distortions. Classifiers may also be fooled by images that look very similar to one another but represent different objects, such as chihuahuas and blueberry muffins, or sheepdogs and mops.

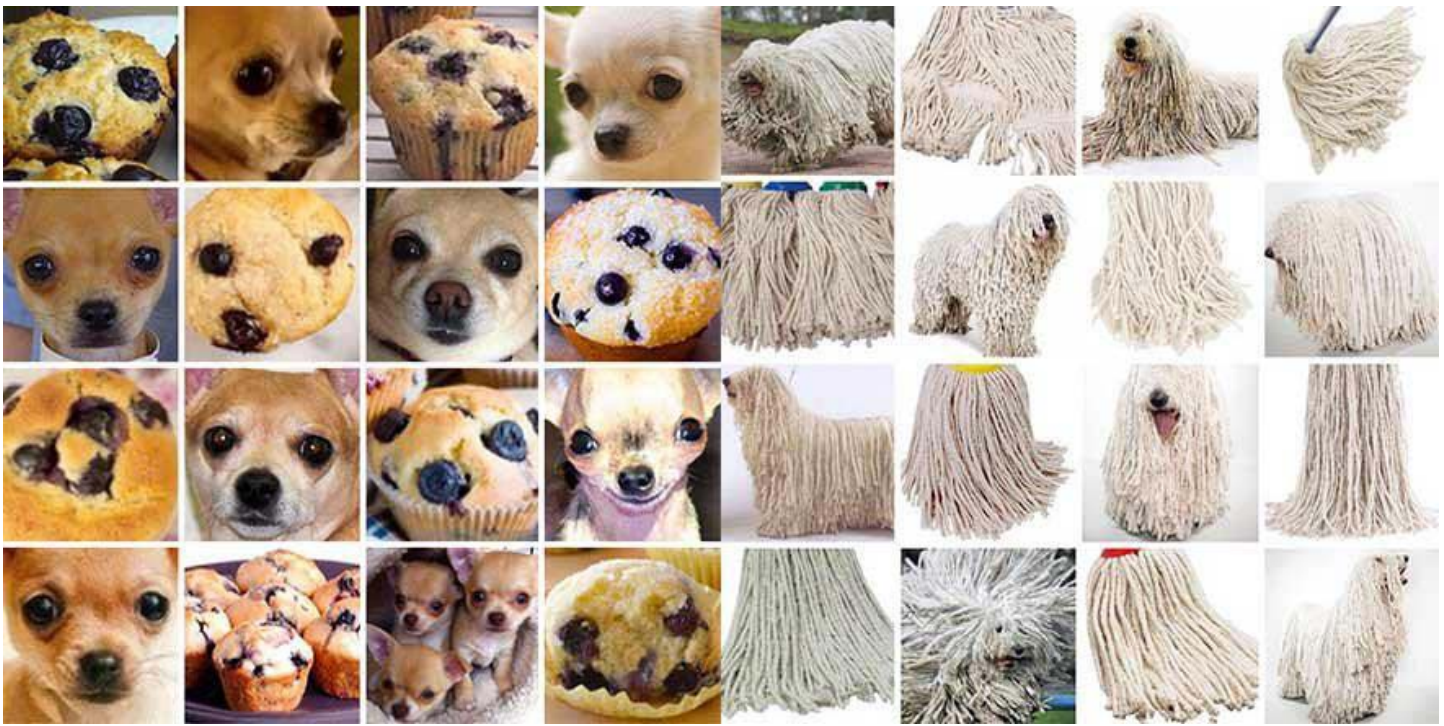

▲ **Figure 1.** Visually similar images: chihuahuas and blueberry muffins, or sheepdogs and mops. Source: https://twitter.com/teenybiscuit/status/707670947830968320 (Accessed March 2021).

**Object detectors**. These tools go beyond classifiers by localizing one or more objects in an image and classifying those objects. The output of a detector is typically a location, denoted by a "bounding box," and the class of the object. Importantly, detectors can come in many forms, and often feature trade-offs depending on the desire for speed (e.g., measured as frames per second, or FPS) or accuracy (often calculated as a form of precision or, as described above, the proportion of all positive predictions that are true positive). For example, the use of lower resolution images can result in higher FPS rates, but lower average precision (Huang et al., 2017).

**Semantic Segmentation and Instance Segmentation.** Segmentation tasks are important for content analysis because they are the building blocks for parsing relationships between objects in images and video. Semantic segmentation seeks to be more granular than object detection, by assigning a class label to each individual pixel in an image. Instance segmentation seeks to be even more precise and identify individual object boundaries.

Note that for instance segmentation, two adjacent dogs are differentiated. In semantic segmentation, these would be the same color and not differentiated.

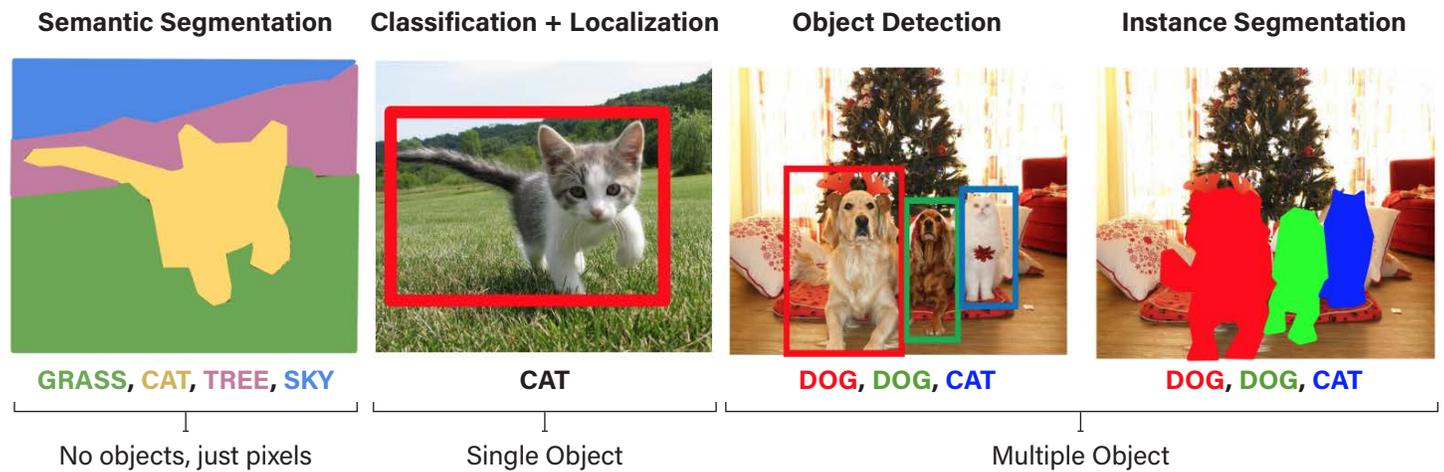

| Semantic Segmentation | Classification + Localization | Object Detection | Instance Segmentation |
| --- | --- | --- | --- |
| GRASS, CAT, TREE, SKY | CAT | DOG, DOG, CAT | DOG, DOG, CAT |
| No objects, just pixels | Single Object | Multiple Object | |

▲ **Figure 2.** Comparing segmentation, classification, and detection. Source: http://cs231n.stanford.edu/slides/2017/cs231n_2017_lecture11.pdf#page=53 (Accessed May 2021).

**Scene understanding.** These tools seek to comprehend a scene by considering the geometric and semantic relationships of its contents. Scene understanding algorithms have important applications in content analysis as they piece together the larger correlations between individual objects. For example, an image containing "fire" might be a campfire or it could be a natural disaster or violent scene. Scene understanding is a compound task that involves a number of the above tasks.

**Object tracking.** This involves following the location of a given object over time in either pre-recorded video or a live stream. Video understanding is a significantly more difficult task than identification of objects in static images because it involves a temporal dimension (i.e., the order of the images matter).

Computer audition focuses on audio content. Some of the techniques used in computer vision are relevant to computer audition, but they are typically conducted on spectrograms (graphic frequency depictions) of audio, rather than on images. Hence, tasks of audio classification are often analogous to their image counterparts. For example, computational auditory scene recognition (CASR) involves predicting the environment in which an audio signal is received. Note that where audio involves humans speaking, speech will often first be transcribed to a text form and integrated with natural language processing (NLP) methods. NLP methods and their strengths and limitations are covered in detail in CDT's previous Mixed Messages report (Duarte et al., 2017).

Modern research in these fields involves the study of **deep learning** models, and specifically **convolutional neural networks (CNNs)**. These models utilize artificial intelligence that is "trained" to learn patterns based on large training datasets (also see the Appendix for a more detailed description of these models and techniques). This implies that the efficacy of these tools are dependent in part on the quality and size of the data sets used for training the model. For example, some forms of content, such as nudity or gore, may have exponentially more publicly available media to train on than "terrorist" content.

## STRENGTHS, WEAKNESSES, AND CONSIDERATIONS OF PREDICTIVE MODELS

The results of individual predictive models are tailored to a variety of specific contexts, and are constantly evolving. This often makes it impractical to make specific claims regarding specific models, especially without adequate insights as to how those models operate and how they arrive at a given output, prediction, or decision. Developing these insights addresses the problem of **explainability**, which is detailed more later.

However, several claims can be made about predictive models generally. Predictive models typically perform better at the "building block" tasks such as classification and object detection. Simpler and more objective questions, such as determining whether an image contains full nudity, blood, or a specific object like a weapon, may see fairly strong performance. Conversely, the more complex a computer perception problem,

▼ **Table 1.** A table outlining the various levels of difficulty associated with particular tasks.

the greater the challenge it presents. Action recognition, such as detecting violence or first-person egocentric action (i.e. GoPro footage), is difficult to deploy in a widespread manner (Kazakos et al., 2019; Perez et al., 2019). A broad classification of various tasks might look as follows:

| Simpler | More Difficult | Very Difficult |
|---|---|---|
| • Identifying whether a given image contains objects (i.e. contraband, symbols); <br> • Identifying objective qualities (blood, clear nudity); <br> • Transcribing speech that is clearly spoken in a common language. | • Differentiating objects that are very similar; <br> • Overcoming attempts to circumvent detection; <br> • Understanding speech in a low-quality recording or with background noise; <br> • Identifying what is happening in a scene. | • Complex action recognition; <br> • Live video analysis; <br> • Understanding subjective context. |

Perhaps the biggest general weakness of predictive models, as with the matching models covered earlier, is that they struggle to be robust, or able to predictably handle changes in inputs that occur either naturally or as a result of circumvention efforts. Object detection and tracking tasks struggle with *occlusion*, or partial or complete covering of one object by another object (Asad et al., 2020). Researchers have tried to develop various methods to improve robustness, though addressing robustness is a bigger problem than simply solving the puzzle embedded in any particular dataset (Cavey et al., 2020; Fawzi et al., 2018). That is akin to memorizing the specific answers for an exam, but failing to understand and apply the actual concepts learned. The challenge as one researcher concluded is that "the human vision system is robust in ways that existing computer vision systems are not" (Hendrycks & Dietterich, 2019, p. 1).

To help provide more objective measures of robustness, Hendrycks presented several benchmarks, adding corruptions and perturbations to the popular ImageNet database, see **https://github.com/hendrycks/robustness**. Accessed March 2021.

For example, computer vision models may not use the same strategy to recognize objects as humans do. Research suggests that humans place more emphasis on an object's shape when it comes to classification, while convolutional neural networks (CNNs) are more likely to rely on texture and color; and that CNNs that learn classification based on shape-based representation may be more robust (Geirhos et al., 2018). These results were echoed by researchers at NVIDIA, who observed, more generally, that gaps exist between how machines attempt to grasp patterns and how humans recognize concepts (Nie et al., 2020). However, these differences are not always obvious, and research between Stanford and Google concluded that "people who build and interact with tools for computer vision, especially those without extensive training in machine learning, often have a mental model of computer vision models as similar to human vision. Our findings contribute to a body of work showing that this view is actually far from correct" (Hendrycks et al., 2020; Hermann et al., 2019, p. 9).

## KEY TAKEAWAYS REGARDING PREDICTIVE MODELS

- Predictive models for multimedia analysis rely on artificial intelligence and learn to recognize patterns in the underlying images, video, or audio that they are trained on. As a result, these models aim to identify *characteristics* and *features* of content.

- Predictive models may perform well in identifying *objective* features in multimedia. Examples may include whether multimedia contains clear nudity, blood, or discrete objects.

- At the same time, predictive models face challenges in considering context. Attempts to capture context are under development. Models are highly dependent on the quality and amount of the data they are trained on.

- Some predictive analysis tasks are considerably more difficult than others. Analysis of static images is much more straightforward than video analysis. Automated real-time video content analysis is highly challenging, both computationally and conceptually. Asking a computer to recognize if an image contains full nudity is thus completely different from asking whether a video depicts a hateful demonstration.

- For different computer perception tasks, various techniques may be available and appropriate depending on specific demands. Some techniques often come with tradeoffs—i.e. expecting faster results may come at the expense of accuracy.

**Predictive models typically perform better at the "building block" tasks such as classification and object detection. Simpler and objective questions, such as determining whether an image contains full nudity, blood, or a specific object like a weapon, may see fairly strong performance. Conversely, the more complex a computer perception problem, the greater the challenge it presents.**

# II. Five Limitations of Tools Used for Automated Content Analysis of Multimedia

While automated tools present some promising use cases when implemented with proper safeguards, their limitations must also be considered in any potential application. This is particularly important in use cases that may have widespread impacts on freedom of expression or user safety when these tools are deployed at scale. We argue that policymakers and developers should understand these limitations when considering what role these tools may play in the analysis of user-generated content.

## A. Robustness

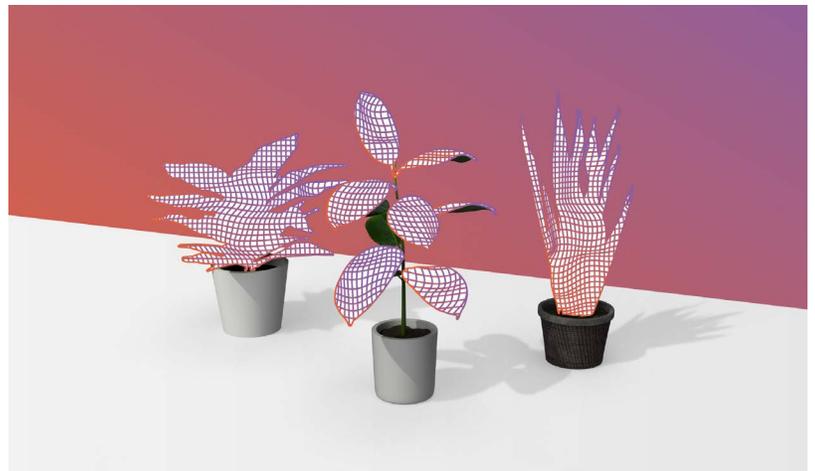

Automated content analysis tools struggle to handle changes in inputs that occur either naturally or as a result of circumvention efforts—in other words, they are not *robust*. "There are no fixes for the fundamental brittleness of deep neural networks," argued one Google AI engineer in 2019 (Heaven, 2019, p. 164). Indeed, the fragility of AI-based prediction systems is well-accepted in the machine learning space. The previous sections of this report have examined the ways in which both matching and predictive models struggle with robustness.

## CIRCUMVENTION EFFORTS

Robustness against circumvention efforts is a recurring problem for multimedia analysis tools. Circumvention efforts do not require inside access to or knowledge of a model to be successful (these are known as "black-box" adversarial attacks). For instance, one 2018 study demonstrated this when they perturbed images to convince the Google Cloud Vision API that skiers in images were dogs (Ilyas et al., 2018). Some of these specific issues in Google's Cloud Vision API have since been mitigated, but the underlying issues in robustness leading up to them remain. Similarly, "adversarial patches" which can be scrambling patterns on clothing or handheld items (the person on the right in Figure 4) have been demonstrated to fool computer vision-based person detection, in this case automated surveillance cameras.

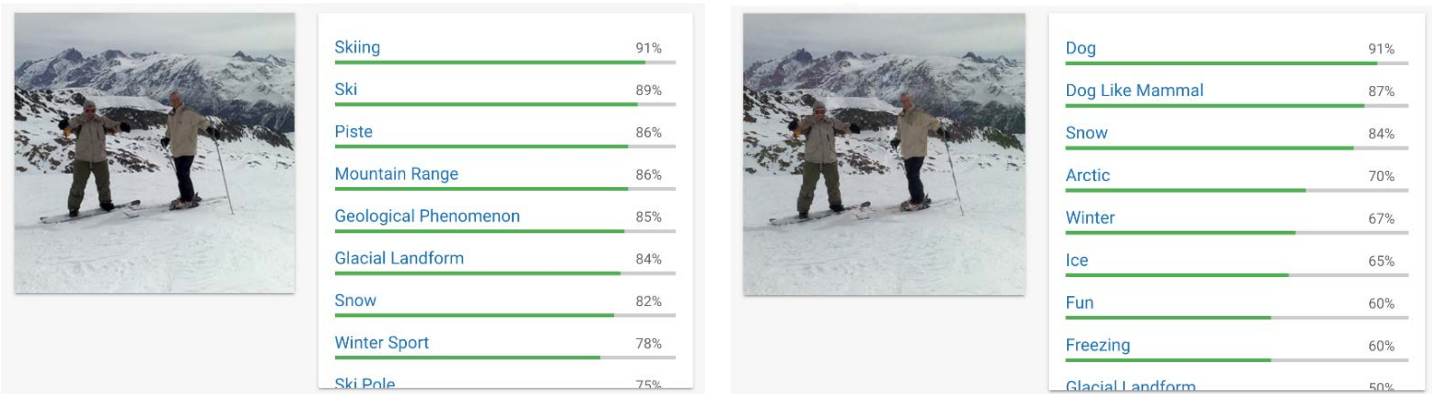

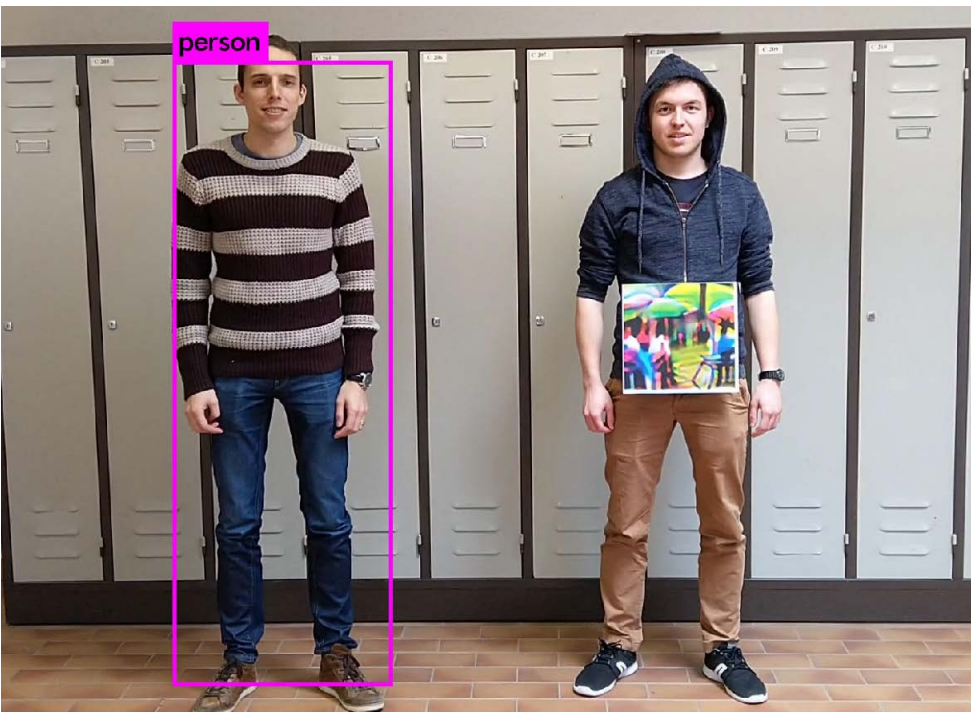

▲ **Figure 3.** Perturbed images implying skiers as dogs. Source: Ilyas, A., et al. (2018).

◀ **Figure 4.** Demonstration of "adversarial patches." Source: Thys, S., et al. (2019).

Similar results have been achieved in safety-critical instances where researchers found that slight real-world perturbations to stop signs could trick a computer into thinking the signs said yield or, worse, speed limits of 45, 70, or 100 (Eykholt et al., 2018). While many of those specific issues have been mitigated, they reveal the stakes of failure (Lu et al., 2017).

Some research has asserted that generating synthetic data is not enough to improve robustness alone. Rather, training on diverse real-world data is what really makes a difference in a model's results because many real-world examples are "adversarial" by nature in ways that may be difficult to duplicate synthetically (Taori et al., 2020). In other words, very tricky real-world examples can easily degrade classifier performance (Hendrycks et al., 2021).

▼ **Figure 5.** Examples of naturally-occurring adversarial images. Source: (Hendrycks et al., 2021).

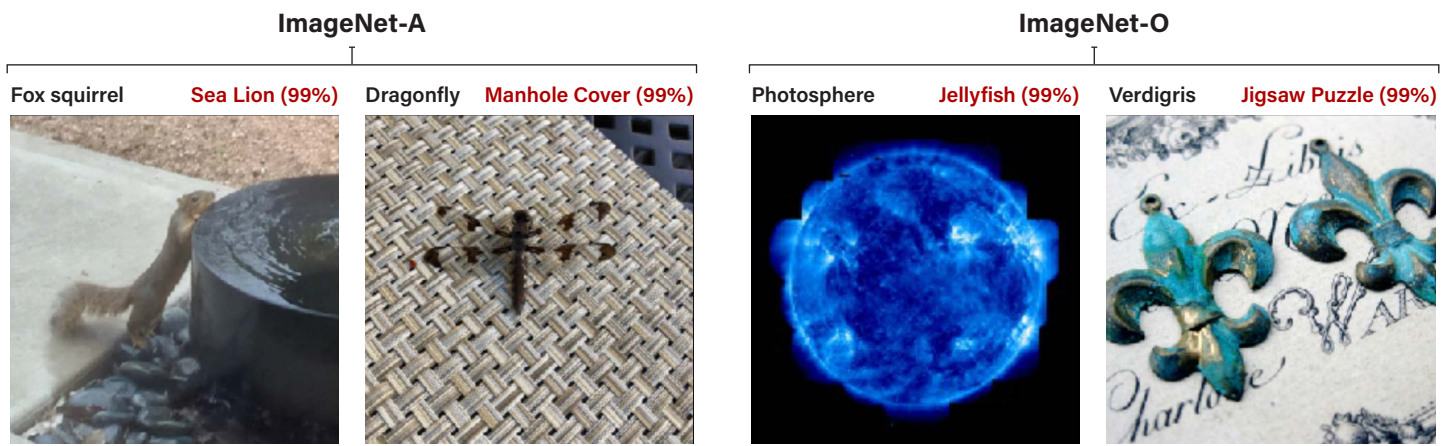

### EVOLVING THREATS SUCH AS DEEPFAKES

Circumvention efforts are constantly evolving, such as deepfakes, which present advanced challenges for automated systems. Even as tools are modified to address problems, such as the example in Figure 3 of Google's Cloud Vision API improving to address the perturbation of images, circumvention efforts are similarly evolving and becoming more sophisticated. One example of this is the case of deepfakes – synthetic manipulations of identities and expressions that may make it appear as if an individual's face is on another's body, speaking or doing things that they never did. For example, in one video, former U.S. President Barack Obama appeared to say numerous expletives in a public address (Mack, 2018). In fact, producer Jordan Peele was projecting his own words onto an AI-animated version of President Obama to warn of the dangers of deepfakes. There are some legitimate use cases of the technologies underlying deepfakes, in fields like movie production, game design, or improving the quality of real-time video streams (Vincent, 2020). But when weaponized, they have the potential to cause serious reputational harm and spread disinformation. Deepfakes have been used, for instance, to project the faces of female celebrities onto pornographic videos, or to perpetuate gender-based violence against non-celebrities (Romano, 2018).

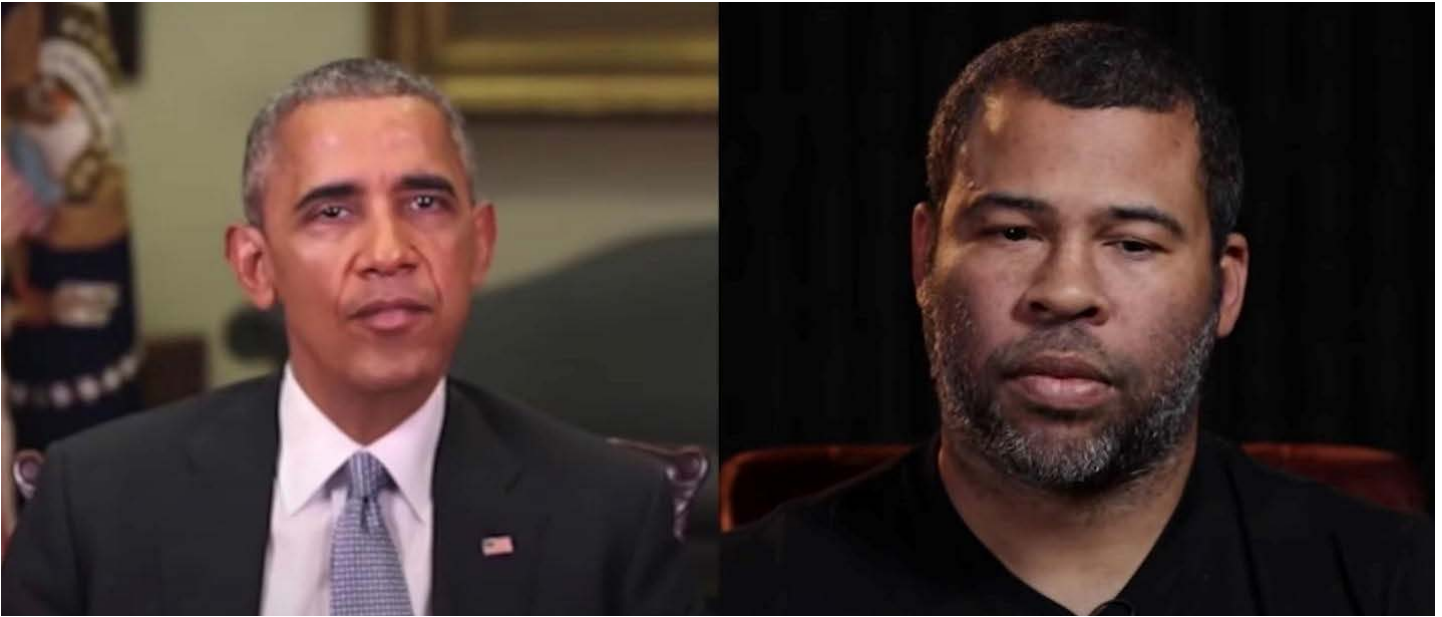

Deepfake detection is presently a major industry priority and challenge. Efforts such as the Deepfake Detection Challenge (DFDC) dataset were created within industry to encourage research into mitigation methods. Some proposals involved extracting visual and temporal features from faces (Montserrat et al., 2020). Another promising method, proposed by the developer of perceptual hashing methods like PhotoDNA and eGlyph, offers a biometric-based forensic technique for detecting face-swap deepfakes. The approach utilizes CNNs to learn facial and head cues, identifying that individuals often communicate not only with their faces but head movements that can be learned by computers (Agarwal et al., 2020). Responses like these continue to try to stay a step ahead of adversarial methods that are constantly adapting.

▲ **Figure 6.** The expressions on the left of former President Barack Obama are actually projections of expressions by Jordan Peele. Source: Mack, D. (2018).

# B. Data Quality

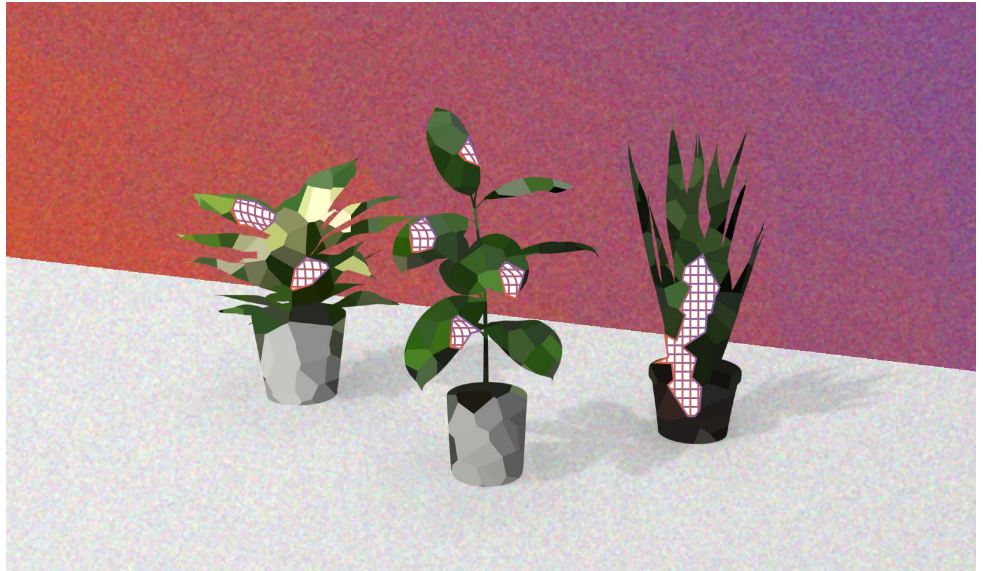

Decisions based on automated multimedia content analysis risk amplifying biases present in the real world; this is often due to poor *data quality*. Training deep neural networks involves exposing them to massive amounts of data. However, numerous steps must be taken to ensure that training data does not serve to amplify biases. Visual and auditory components of multimedia create even more opportunities for bias than those present in text.

### DATASETS PROPAGATE UNDERLYING BIASES

It is well-accepted that datasets have real biases. As one expert observed, datasets are "a look in the mirror . . . they reflect the inequalities of our society" (Condliffe, 2019). Biases manifest in multimedia analysis tools. Amazon Rekognition's facial recognition technology, on a test conducted by the ACLU, mistakenly matched 28 members of Congress with a mugshot database and identified them as having been arrested for crimes (Snow, 2018). More importantly, the false matches disproportionately included Congressional members of color, including the late civil rights legend Rep. John Lewis (D-Ga.). Twitter's AI-generated photo previews faced scrutiny when they appeared to favor white faces over those of persons of color (Lyons, 2020). And reporting by the *Washington Post* found that "smart speakers" such as Alexa or Google Assistant showed significant disparities in understanding people with different regional accents in the United States (Harwell, 2018). AI systems do not learn biases in a vacuum, and issues are quite often traced back to deficiencies in training data.

Once biases are encoded, they are often hard to detect. Researchers have proposed classifier benchmarks that are demographically and phenotypically balanced, as well as deliberate metrics for subgroups such as "darker females," "darker males," "lighter females," and "lighter males" (Buolamwini & Gebru, 2018).

The causes of bias in data are numerous. As Microsoft has noted, "Most real datasets have hidden biases" (Microsoft, n.d.). Though beyond the scope of this paper, we note the current research into novel bias mitigation methods, such as rethinking crowdsourced microtasking services like Mechanical Turk (MTurk) which are often relied upon to provide data labeling (Barbosa & Chen, 2019).

But even such efforts may adopt binary views of race or gender, and contribute to trans erasure (West et al., 2019). Efforts like the Inclusive Images dataset attempt to correct for "amerocentric and eurocentric" biases present in popular datasets like ImageNet and Open Images by including more balanced representation of images from countries in Africa, Asia, and South America (Shankar et al., 2017). The nature of photography itself may mean prediction tools always struggle with racial bias, Zoé Samudzi argues, due to "color films' historical non-registry of black people"(Samudzi, 2019). Potential biases in data sets are not just limited to skin color but also virtually any other characteristic. For instance, a content analysis tool that is only trained on western weddings may associate "wedding dress" with western presentation of what a bride looks like (i.e., a white formal gown), instead of non-western wedding dresses (Wiggers, 2018).

For more details on the types of biases found in ImageNet see Parbhu and Birhane (2020) and Steed and Caliskan (2021).

## INSUFFICIENT DATA AVAILABLE IN THE REAL WORLD

Some categories of content do not have sufficient data available in the real world and present training data problems by their very nature. There is much more multimedia content of nudity than of gun-based violence because "thankfully, we don't have a lot of examples of real people shooting other people," said Yann LeCun, Facebook's chief AI scientist (Kahn, 2019). So-called "terrorist" content presents another challenge. Despite prominent coverage in the news media, from a data science perspective there simply is not much publicly available multimedia from designated terrorist organizations to train models on (though there are recent emerging efforts to attempt to mitigate this). This is a serious and recurring problem in data science known as class imbalance, where some classes have a significantly higher number of examples in the training set than others (Kushwaha, 2019). These issues can exist at a global scale, where lack of data may reflect variances in digital footprints around the world, or they can exist locally, based on relative access to technology within communities. Yuille and Liu argue that data in the real world is combinatorially large, and it is difficult for any dataset, no matter how large, to truly represent the complexity of real life (Yuille & Liu, 2019).

The Terrorist Content Analytics Platform (TCAP) is an effort by a key partner of GIFCT to produce a "transparency-centred terrorist content tool." **https://www.terrorismanalytics.org/**. As a stated goal, it seeks to include civil society in all stages in the process. Tech Against Terrorism (2020).

## DATA-LEVEL BIAS MITIGATION ONLY GOES SO FAR

Some strategies to address potential bias in datasets (for example because of class imbalance) attempt to oversample to replicate and synthesize new samples from an underrepresented class of data types (Buda et al., 2018). Google released a framework called "MinDiff" to attempt to mitigate bias by optimizing a sample of data based on certain fairness metrics (Prost et al., 2019).

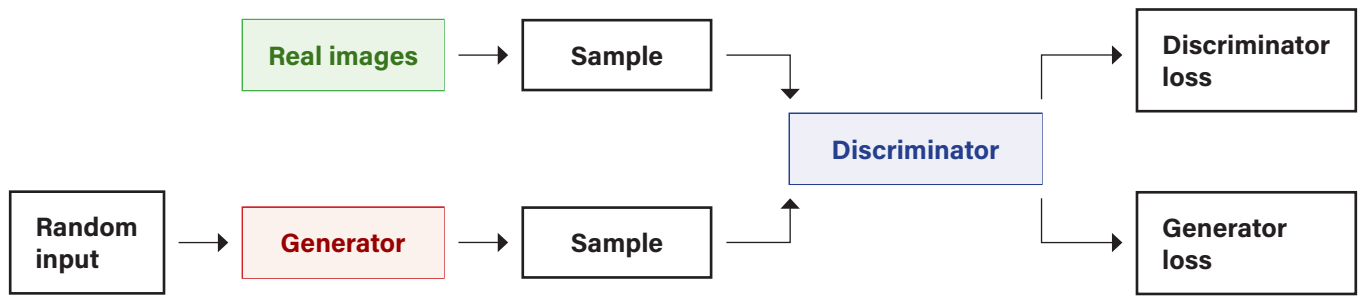

▲ **Figure 7.** An illustration of GAN. Source: https://developers.google.com/machine-learning/gan/gan_structure.

Tools known as generative adversarial networks (GANs – see Appendix for a more detailed explanation) may be used to generate synthetic data, or data that can mimic real world data based on certain parameters (Goodfellow et al., 2014). GANs consist of two competing neural networks, akin to a boxing match between a *generator* and a *discriminator* that are both improving at their respective task by checking one another. An example of a GAN would involve a generator network creating fake images, which then are fed into the discriminator network which must determine, based on its training off real images, whether or not the images it is fed are fake. Those labels are then compared to the "ground truth." The discriminator learns from its mistakes, while the generator adapts based on what successfully fooled the discriminator to present more sophisticated fakes. Two respective feedback loops allow for both models to improve. The result in this case is a synthetic data set that can be used, for example, to address a prior lack of real world data (Q. Wang et al., 2019) or to improve the accuracy of a predictive model when applied from one context to another (Sankaranarayanan et al., 2018).

However, training on synthetic data may present risks of overfitting, or conforming too closely to one training set in a way that does not generalize well. This is because the models used to generate the synthetic data are vulnerable to the same limitations described here, including a lack of robustness, lack of context, etc. Thus, even actively correcting for biases in the training data may be insufficient. For example, West, Whittaker, and Crawford of the AI Now Institute warn that "[t]he histories of 'race science' are a grim reminder that race and gender classification based on appearance is scientifically flawed and easily abused" (West et al., 2019, p. 3). Similarly, Julia Powles calls attempts to address bias a distraction. "Bias is a social problem, and seeking to solve it within the logic of automation is always going to be inadequate" (Powles, 2018).

**It is well-accepted that datasets have real biases. As one expert observed, datasets are "a look in the mirror . . . they reflect the inequalities of our society" (Condliffe, 2019).**

# C. Lack of Context

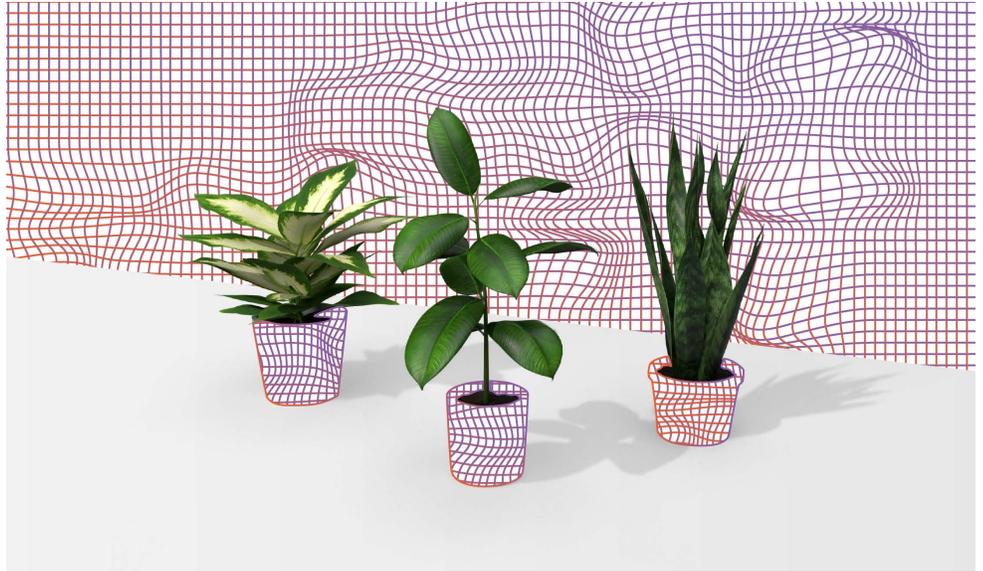

**Machine learning tools often struggle to appreciate the same contexts that humans can readily recognize. Tasks that may seem simple for a human may often be highly context-dependent and thus present significant challenges for algorithmic models.**

Automated tools perform poorly when tasked with decisions requiring judgment or appreciation of cultural, linguistic, social, historical, or other *context*. These tools often struggle to appreciate the same contexts that humans can readily recognize. Tasks that may seem simple for a human may often be highly context-dependent and thus present significant challenges for algorithmic models. For example, TikTok videos showing an adult kissing a young child could be incredibly harmful, or completely benign (as in the case of a parent tucking a child into bed). Similarly, images or videos of the identical activities in summer or winter may be treated differently by a model purely because people may wear fewer clothes in warm weather.

As described earlier, matching models can at best identify two pieces of content (e.g., two images) that are identical (cryptographic hashing) or sufficiently identical (perceptual hashing) but not the context in which either is used. Similarly, prediction models are currently more accurate when it comes to image classifiers and object detectors than more compound tasks, such as scene understanding where context is relevant. Thus, for example, they may be able to identify nudity, but not make a judgment about whether that nudity is occurring in the context of artistic expression or abuse. In general, the current approaches that matching and predictive models use to identify patterns differ in how humans recognize concepts, particularly with regard to the importance of context.

Companies are stepping up efforts to incorporate context in analysis tools, but many of these holistic methods are still in their early stages and face many challenges. Facebook is currently researching identification of hateful memes, where images and text that may seem benign in isolation may cause harm when combined in a meme (DrivenData, 2020). Facebook also implements tools called Whole Post Integrity Embeddings (WPIE) to understand content across modalities. It has utilized WPIE to ascertain context to determine when user posts and accompanying photos are attempting to sell illicit drugs (Facebook, 2019). These algorithms attempt to parse whether a "full batch" of "special treats" accompanied by a picture of baked goods is talking about Rice Krispies squares or edibles.

On Twitch, automated methods may be used to produce highlight reels of noteworthy moments in video game streams. One method proposed *jointly* considering emotion detection of a streamer's facial expressions, game scene analysis, and audio stream analysis (Ringer & Nicolaou, 2018). Violence detection has seen recent research in weakly-supervised multimodal methods utilizing audio cues (Wu et al., 2020). Some research has proposed models to identify broadcasters of adult content on social live streaming services by combining image recognition with follower characteristics, username, and other characteristics (Lykousas et al., 2018). While these techniques are being researched, they are just scratching the surface of genuine context-appreciation by machines.

# D. Measurability

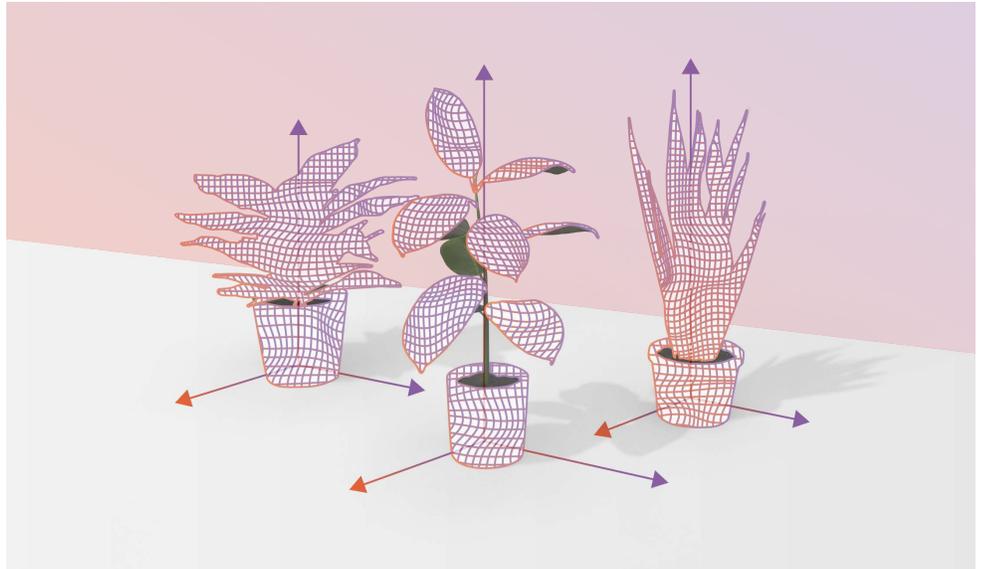

**"Accuracy" in isolation is generally an unhelpful metric. In most instances it is meaningless.**

Generalized claims of accuracy typically do not represent the actual multitude of metrics for model performance; there are many ways to approach *measurement* of the performance of automated systems. However, "accuracy" in isolation is generally an unhelpful metric. In most instances it is meaningless. There are several reasons for this. First, the degree of "accuracy" may be a function of the class imbalance problem, wherein some forms of harmful content are sparse by nature. Therefore, a predictive model that simply says everything is "not terrorist propaganda" will be accurate 99.9% or more of the time. But it would be right *for entirely the wrong reasons*. Second, even if a model predicts the right result 999 out of 1000 times, that one wrong result can have extremely harmful impacts. This is particularly the case when wrong results have high stakes for either freedom of expression or safety. Third, metrics of positive model performance may also be self-selective. Better metrics for measuring predictive models include their precision and recall, as discussed earlier. However, use of any metrics, including *precision* and *recall*, nonetheless run into challenges with sparse data.

Another important consideration in content analysis is the sheer scale of content. 99.9% performance may actually be *quite* bad if the 0.1% means that tens or hundreds of thousands of pieces of user content are false flagged or acted upon incorrectly at scale. These scales raise the stakes of user impact and implications for freedom of expression (Spoerri, 2019). Indeed, e-mail service providers consider any false positive rate higher than approximately 0.1% too high in the use case of spam filters, due to possible limitations on speech.

One measure used to compare various models are standardized datasets, called benchmarks. For example, computer vision researchers may lack a common data set to test different models for particular forms of robustness. Thus, researchers can create benchmarks by taking a commonly used data set and adding a standard set of changes or corruptions, to allow others to test different models for robustness using the modified data set (see for example Mu & Gilmer, 2019). But benchmarks, too, should be scrutinized, and performance on them may highlight specific strengths that may not translate to the real world (see for example Northcutt et al., 2021).

# E. Explainability

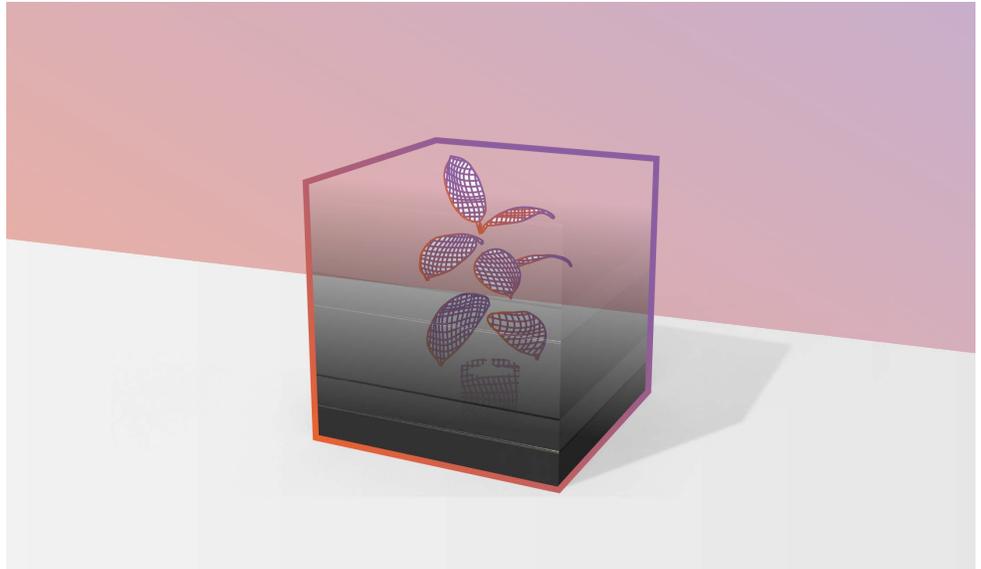

**The "black-box" nature of AI systems is compounded in content analysis, particularly moderation, because moderation itself has "long been a famously opaque and secretive process," as one researcher notes (Gorwa et al., 2020).**

It is often difficult to understand the steps automated tools take in reaching conclusions; many types of machine learning techniques resist easy *explainability*. Some of the largest neural networks used by industry leaders utilize billions of interrelated parameters (Ray, 2020). Neural networks are complex and non-linear, and do not necessarily "show their work," which makes it very difficult to understand how they operate, what features they use to make decisions, and how various decisions are weighted and why (Eilertsen et al., 2020). The "black-box" nature of AI systems is compounded in content analysis, particularly moderation, because moderation itself has "long been a famously opaque and secretive process," as one researcher notes (Gorwa et al., 2020). The lack of transparency in automated decision-making can be exacerbated when commercial intellectual property rights are claimed as a barrier to disclosure. Further, it may become more difficult to ascertain the potential human rights harms of content takedowns where initial flagging decisions are made by automated systems that lack transparency and clarity in the reasons for the takedowns (Gorwa et al., 2020).

## TOOLS AND TECHNIQUES FOR EXPLAINABLE AI

Explainability tools seek to illuminate why a particular algorithm reached the conclusion it did. These tools operate in a variety of ways, but generally attempt to highlight key *features* that were weighted as part of a specific output. For example, an explainability tool for an image classifier may highlight ears, whiskers, and a tail as elements that contributed to a conclusion that an image contains a cat. One approach to do this employs heatmaps to display key regions of an image supporting classification predictions (Karlinsky et al., 2020).

Knowing why a model reaches predictions is important because it may reach the right prediction but for the wrong reasons (Samek, 2020). In one study, the researcher discovered that an algorithm to predict a person's age from a photo learned, for whatever reason, to correlate age with not smiling. In other words, the model believed that the elderly do not laugh (Lapuschkin et al., 2017). These mistakes are difficult to catch without explainability and auditing. Several current techniques exist for getting an AI system to produce cues as to why it produced certain outputs.

These include perturbation techniques (modifying features to test their importance for achieving a result), surrogate/sampling methods (involving approximating predictions locally), and structural techniques (analyzing the inner structure of the network) (Samek, 2020). Numerous industry players have produced tools to interpret AI inferences, including Facebook's Captum, which is open source, and Fiddler's Explainable AI. IBM has an AI Explainability 360 toolkit, which is also open source, and Microsoft released InterpretML.

## DIFFERENT EXPLAINABILITY APPROACHES FOR DIFFERENT STAKEHOLDERS

Explainability may mean different things in different contexts. The use case for a developer, who may need to understand the *structure* or *limitations* of an algorithm, generally is different from that of an end user, who may wish to simply understand why a specific image was analyzed in a particular way. Thus, there is no "one-size-fits-all" approach to explainability (Hind, 2019), and different settings may require different types of explanations.

One proposal by IBM attempts to articulate what these different forms of explainability could look like (Arya et al., 2019). Their taxonomy for explainability includes:

- Directly interpretable explanations (a simple model with decision rules that can be understood by the user) and post hoc interpretable explanations (e.g., a CNN) that require an associated model to provide explanations.
- Global explanations that cover how the entire model operates, and local explanations that explain a specific prediction/output.
- Static explanations that do not change, and interactive explanations that provide more depth or explanations depending on new user requests (e.g., via a dialogue).

In general, explainability remains a nascent research subject and there is much more work to be done. The National Institute for Standards and Technology (NIST) has proposed draft key principles to guide the development of explainability (Phillips et al., 2020). These include: (1) providing an explanation (i.e., evidence or a reason) for the output, (2) making sure that the explanation is understandable to each user (thus this may require providing different reasons for different users), (3) ensuring that the explanation is accurate in describing how the model arrived at a given output (which is different from model precision described earlier), and (4) outlining the limits of its use or cases that the model was not designed for.

# III. Conclusion: Capabilities, Limitations, and Risks

**Some of these limits might be addressed by future technological advances. But many of the limitations are inherent to the technologies themselves. Social networking services, civil society, governments, and others need consider these limitations.**

The Internet has ushered in an explosion of digital speech and communication in the form of text, images, video, and audio created and uploaded by people across the globe. The scale of this content overwhelms our capacities for evaluation, and since the earliest days of the Internet, different forms of automation have been employed to help us filter, sort, rank, and otherwise analyze user-generated content. Advances in machine learning techniques have raised the prospect of much more sophisticated analysis, where "artificial intelligence" grows to approximate, or even surpass, human understanding of the meaning and context of all of this content.

Today's machine learning techniques for analyzing content at scale represent significant technological advancements – and, as discussed earlier, tools employing these techniques can be useful in a variety of scenarios, including content moderation, research, fraud detection, improving accessibility of media files, and more. But these techniques also possess real limitations that affect their utility for different tasks. They struggle to parse context and extract nuanced meaning, and are often vulnerable to evasion and circumvention.

Some of these limits might be addressed by future technological advances. But many of the limitations are inherent to the technologies themselves. Social networking services, civil society, governments, and others need consider these limitations (e.g., robustness, data quality, lack of context, measurability, and explainability) when considering whether and how to use automated tools to address the complex problems of large-scale content analysis.

A failure to address these limitations in the design and implementation of these tools will lead to detrimental impacts on the rights of people affected by automated analysis and decision making. For example, a tool with limited robustness can be easy to circumvent and can fail to identify abusive content, including "deepfakes" or other manipulated media (Chesney & Citron, 2019). Poor data quality can lead to machine learning models that perpetuate existing biases in society (Buolamwini & Gebru, 2018) and that yield outputs with a disparate impact across different demographics, exacerbating inequality (Burton-Harris & Mayor, 2020). Insufficient understanding of context can lead to overbroad limits on speech and wholly inaccurate labeling of speakers as violent, criminal, or abusive (Llansó, 2020a). Poor measures of the accuracy of automated techniques can lead to a flawed understanding of their effectiveness and use, which can lead to over-reliance on automation and inhibit the introduction of necessary safeguards. Finally, limited explainability can restrict the options for remedying both individual errors and systematic issues, which is particularly important where these tools are part of key decision making systems.

Large scale use of the automated content analysis tools described in this paper will only amplify their limitations and associated risks. Discussions of the potential benefits of automated content analysis should not, therefore, understate the critical role of human review or the importance of structuring systems with opportunities for intervention (Duarte et al., 2017). Technology companies that create and employ tools for automated content analysis should include opportunities for human review and intervention throughout the design and implementation processes, as part of a set of safeguards to identify and mitigate adverse human rights impacts of their technology.

**Automated content filtering should never be required by law.**

Policymakers must also take into account the limitations of automated analysis tools before promulgating laws or regulations that require or assume their use. For example, laws that impose extensive content moderation obligations on social media platforms that handle millions of pieces of content on a daily basis may explicitly or implicitly rely on the assumption that automated content analysis tools will make that possible. But that assumption carries with it all the limitations of those tools, which may result in errors and harms that should be weighed in the policymaking process. Indeed, automated content filtering should never be required by law. Policymakers should not embrace or normalize an uncritical view of the efficacy of these tools (Gorwa et al., 2020). This can undermine important and needed public dialogue about what problems machine learning or "artificial intelligence" can – and cannot – help us solve.

# IV. Appendix: Automated Multimedia Content Analysis Techniques

## Matching Models - Cryptographic Hashing

Cryptographic hashing functions create a string of numbers and letters called a hash, which almost uniquely identifies a file. Similar cryptographic functions are also used to encrypt data in applications like e-mail, Signal or WhatsApp texts, or certain file storage mediums, and are meant to assure recipients of the authenticity of a message or file, down to the last bit. Cryptographic hashing uses a cryptographic function to generate a random hash fingerprint. The cryptographic component makes these functions generally "non-smooth" and extremely sensitive to change. This means even miniscule alterations in the input data will drastically change the resulting hash. For example, changing the shade of one pixel in a high-resolution photo would produce a distinct cryptographic hash. Cryptographic functions are also highly collision-resistant, meaning different pieces of content will produce very different hashes so the likelihood of two different pieces of content producing the same hash (or "colliding") are incredibly low (Engstrom & Feamster, 2017).

▼ **Figure 8.** An example of how small changes in input data can lead to very different results in cryptographic hashing. This graphic has been recreated, and based on one by Rosenbaum, K. (2017, June 26). Cryptographic Hashes and Bitcoin, Grokking Bitcoin, Manning Publications. Retrieved December 17, 2020 from https://freecontent.manning.com/cryptographic-hashes-and-bitcoin/.

**Original cat image**

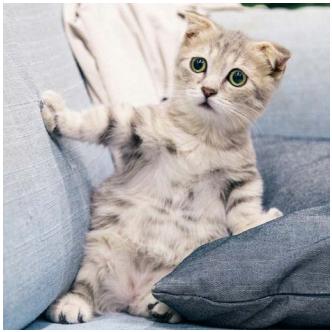

One way MD5 hash generator

MD5 checksum

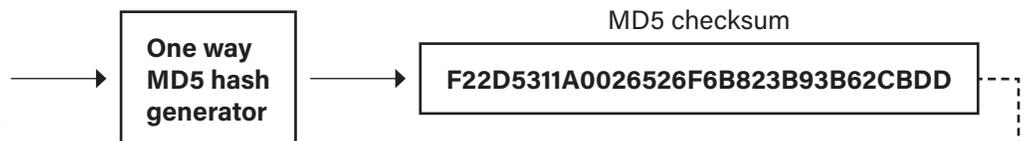

F22D5311A0026526F6B823B93B62CBDD

Completely different resulting hashes

**Modified with blue eyes**

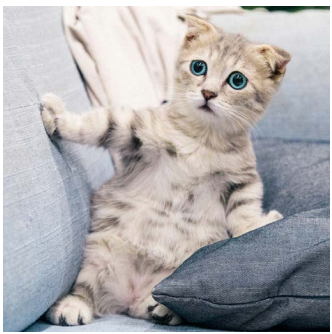

One way MD5 hash generator

MD5 checksum

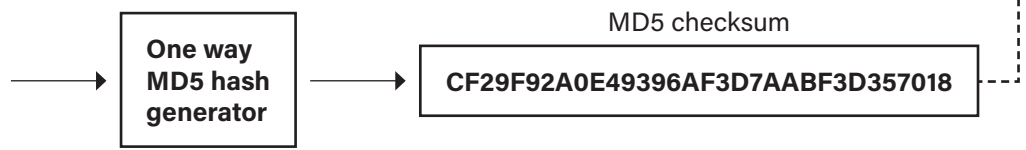

CF29F92A0E49396AF3D7AABF3D357018

Cryptographic hashing is highly effective in authenticating known content without alterations. This leads to its primary drawback in its use for automated content analysis, which is its lack of robustness, meaning it is not resistant to minor data distortion. Substantially identical pieces of content may hash very differently. This is particularly problematic in use cases that are adversarial in nature—i.e. an attacker tries to circumvent a hash-based filter and modifies content such that it produces a different hash. Alterations might also occur naturally, simply through the routine transfer of data which may utilize compression to save bandwidth and space. Most modern content sharing systems apply some form of post-processing which would, by nature, change the bits of the file and thus the output of the cryptographic hash. Ideally, a matching system would be *input invariant*, which means that small alterations in input would produce little or no change in the hash.

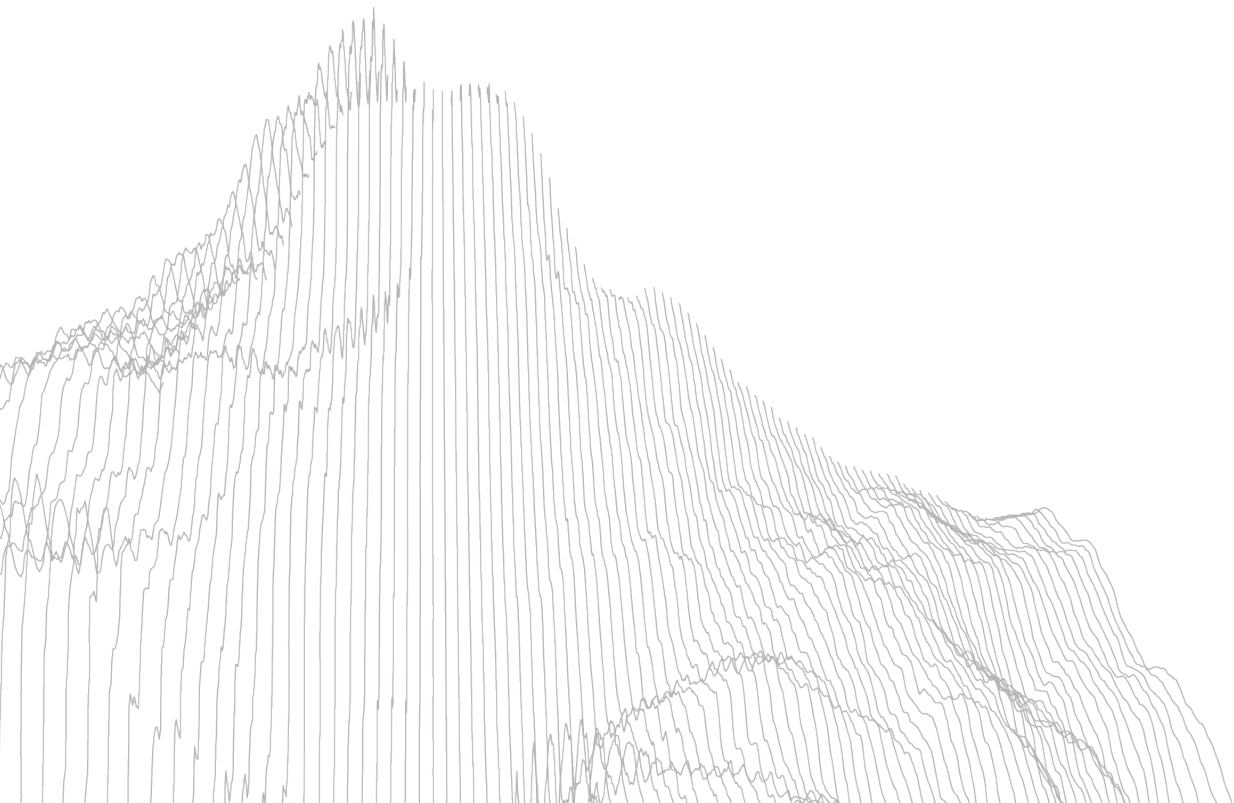

# Matching Models - Perceptual Hashing

Perceptual hashing seeks to determine not whether two pieces of content are identical, but whether they are "alike enough"—i.e. practically identical. For example, if you were shown two photos of the same person, except one single hair in one photo were a slightly different shade, you would likely not notice this miniscule change and consider the photos the same. However, a cryptographic hashing method would consider these completely different. The goal of a perceptual hashing method would be to recognize these as fundamentally the same photo. Perceptual hashing methods aim to better comprehend the nature of a piece of content so that the machine cannot be fooled by imperceptible or non-meaningful changes, such as rotations, resizing, orientation flips, noise, delays in audio or video, or watermarking. Some of these changes might be naturally occurring, or others may be human-designed efforts to circumvent detection.

Some of those methods include ones based on invariant features, local feature points, dimension reduction, and statistics features. (Du et al., 2020).

Perceptual hashing methods involve various methods of pre-processing content, hashing it, and using metrics to compare how alike two pieces of hashed content are. A threshold can be set to determine what degree of difference between hashes is allowed to still consider them matches. Modern perceptual hashing methods apply a range of techniques, including different approaches to create hash fingerprints. For example, by applying a grid and analyzing relationships among pixels in each square, the hash comparison is able to recognize the underlying similarity of the images (see Fig 9).

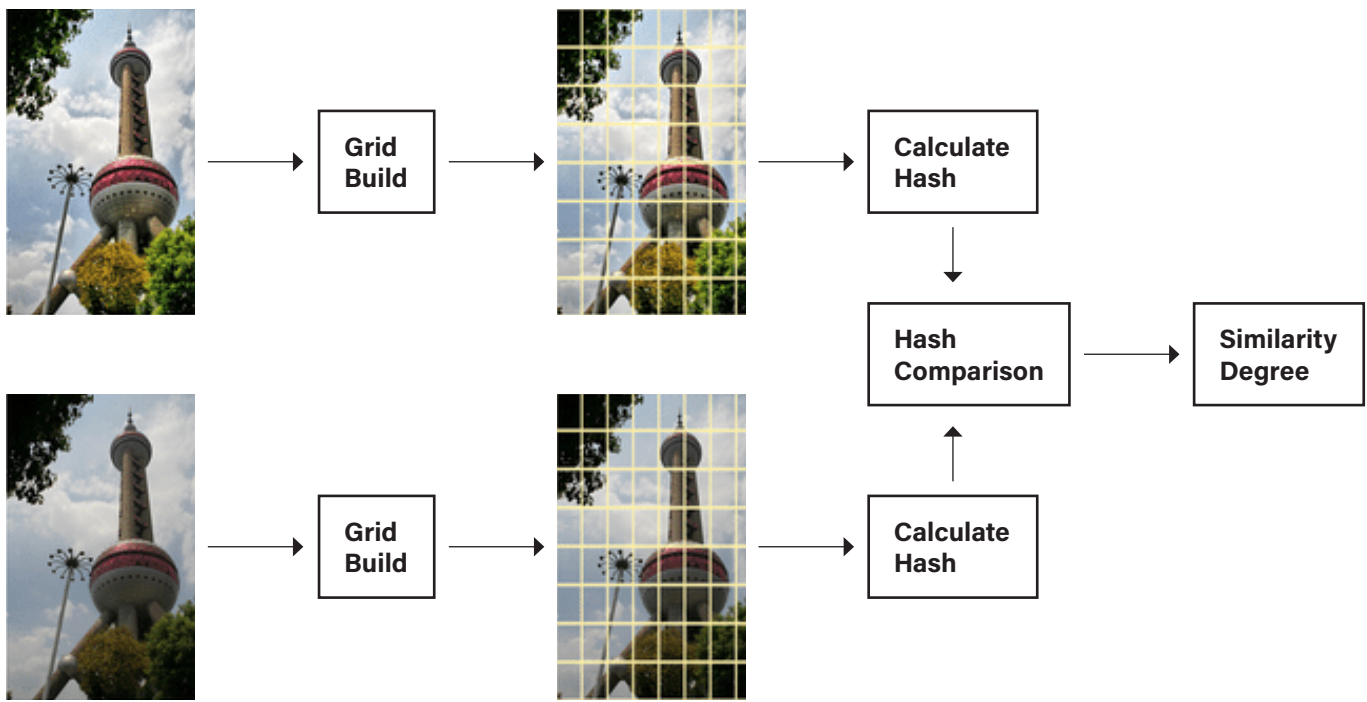

▲ **Figure 9.** An overview of a hash comparison process of two versions of the same photo but with different levels of color saturation. Source: Souza et al. (2018). https://www.researchgate.net/profile/Veronica-Teichrieb/publication/325521472_Generating_an_Album_with_the_Best_Media_Using_Computer_Vision/links/5b2179a6458515270fc6da3e/Generating-an-Album-with-the-Best-Media-Using-Computer-Vision.pdf.

Perceptual hashing methods offer more flexibility than their cryptographic counterparts. For instance, they can be capable of identifying content that is hidden within other pieces of content, such as a video that is masked within another video (Langston, 2018). In order to evolve as attackers evolve, perceptual hashing functions may utilize techniques like deep learning and convolutional neural networks (discussed in more detail in Box 1) in order to adaptively identify manipulation methods and features. Such methods have shown promise, with the ability to distinguish between substantively distinct images, while also not being fooled by superficial changes (Jiang & Pang, 2018). Some specific implementations of perceptual hash algorithms include systems designed to detect child sexual abuse material (CSAM), terrorist propoganda, and copyrighted content.

## CHILD SEXUAL ABUSE MATERIAL (CSAM)

Perceptual hashing has been the primary technology utilized to mitigate the spread of CSAM, since the same materials are often repeatedly shared, and databases of offending content are maintained by institutions like the National Center for Missing and Exploited Children (NCMEC) and its international analogue, the International Centre for Missing & Exploited Children (ICMEC) (Lee et al., 2020). PhotoDNA, developed by Microsoft, is presently the most widespread perceptual matching method for countering CSAM. At a high level, it works by first converting a full-resolution color image to grayscale, then downsizing it to 400 x 400 pixels. A filter is applied, the image is partitioned, and then measurements are extracted onto feature vectors which are compared using a distance metric. PhotoDNA for video applies a similar method to certain video "key frames" (Langston, 2018). More specific information about the PhotoDNA algorithm and the NCMEC database are not publicly available, due to concerns that attackers would use that information to circumvent these protections; however, this lack of transparency also closes off avenues for independent audits and review.

Facebook has open-sourced its PDQ and TMK+PDQF algorithms for image- and video-matching, respectively (Davis & Rosen, 2019). PDQ, based on an algorithm called pHash, stores and compares the outputs of 16 x 16 transformations of images. Other perceptual applications in CSAM include CSAI Match, a proprietary hash-matching technology developed by YouTube, which is utilized by Adobe, Tumblr, and Reddit. Google released an open-source Content Safety API, an AI-powered tool grading the severity of disturbing images, with the Internet Watch Foundation (Todorovic & Chaudhuri, 2018). New methods propose purely metadata-based analysis (meaning they work without examining the actual content of a file) using file paths, which could augment perceptual hashing methods in the fight against CSAM (Pereira et al., 2020). In practice, companies may use a combination of these and other automated tools to detect CSAM on their networks.

See for example the tools used by Pornhub to detect CSAM and non-consensual content. https://help.pornhub.com/hc/en-us/articles/1260803955549-Transparency-Report/. Accessed April 2021.

### TERRORIST PROPAGANDA

The Global Internet Forum to Counter Terrorism (GIFCT), a consortium founded by Facebook, Microsoft, Twitter, and YouTube and now operating as an independent entity, maintains a shared industry hash database of what they view as terrorist and violent extremist content. Individual companies that are members of the consortium may, depending on the nature of their participation, contribute content they deem to include terrorist propaganda to be catalogued in the shared database. This shared database is not available for independent review or audit.

Joining the consortium involves signing an NDA, MOU, and obtaining licenses to use hashing techniques. As a result, the technical workings of the SIHD are not publicly known. See **https://gifct.org/**. Accessed March 2021.

eGLYPH is another hashing algorithm for terrorist content created by Hany Farid, the same researcher who developed the original PhotoDNA technology. eGLYPH operates very similarly to PhotoDNA, involving the grayscale conversion of images and down-sizing to 400 x 400 fixed resolution. The algorithm can be used to find videos as well, by filtering out redundant frames to reduce length and file size, and then producing arbitrary-length hashes which can be compared using a "longest common substring." Longest common substring is a technique for comparing how similar two alphanumeric strings are by finding the longest contiguous stretch the two strings have in common. For example, consider the random strings "982tiu3hhuiuh" and "293rr928iu3hhu2tiu." These strings have the common substrings "982," "2tiu," and "iu3hhu." Because "iu3hhu" is the longest of these substrings at six characters long, that is the strongest point of similarity and thus the string used to score the strength of similarity between the two longer strings. This approach can also be used to compare audio files (Counter Extremism Project, 2018; Greenemeier, 2017).

### COPYRIGHTED CONTENT

Copyright-enforcement tools seek to match user-uploaded content to instances of known, copyrighted content. Perceptual methods are often useful in these efforts, since pirated content might add modifications or watermarks to avoid identification. An example of one tool is the Echoprint API, an open-source fingerprinting library utilized by Echo Nest, a subsidiary of Spotify (Ellis & Whitman, 2013). Echoprint contains three components: 1) a code/fingerprint generator; 2) a query server that stores codes to match against; and 3) codes themselves that are used to match against the fingerprints of any given audio files. Specifically, Echoprint creates time/hash pairs based on relative timing between beat-like onsets, and identifies pieces of audio via these pairs. The fingerprint is based on the relative locations of these onsets (*Welcome to Echoprint*, n.d.).

Another example of a similar fingerprinting technology is YouTube's Content ID, which allows rights holders themselves to create fingerprints of their multimedia (Engstrom & Feamster, 2017). The company Audible Magic produces matching systems utilized by major entertainment studios (Universal Music Group, Warner Bros., Sony, and Disney), as well as platforms such as Facebook, Soundcloud, Twitch, and Tumblr. Audible Magic holds numerous patents in perceptual fingerprinting and automated content recognition, including methods for creating unique audio signatures via segmentation. While its methods are proprietary, those patents indicate that it utilizes principles analogous to segmentation and fingerprinting of spectrograms (visual representations of a spectrum of frequencies in a piece of audio).

## OTHER APPLICATIONS

Matching algorithms may appear in any case where an organization wants to blocklist content and flag that content when it appears. For instance, online social matchmaking services like OkCupid have utilized perceptual hashing algorithms to scan for re-uploads of banned profiles (Jablons, 2017). Facebook, too, utilizes a large-scale matching infrastructure called SimSearchNet/SimSearchNet++ on "every image uploaded to Instagram and Facebook" to scan against an existing curated database of "misinformation," including COVID-19 misinformation (Facebook, 2020). Amazon utilizes audio fingerprinting to prevent mentions of the word "Alexa" in advertisements from mistakenly triggering Alexa devices and resulting in negative customer experiences (Rodehorst, 2019).

Content ID was originally licensed by YouTube in 2006 from Audible Magic; after Google acquired YouTube, it acquired a trademark for "Content ID," after which Audible Magic sued Google over use of the term (Sanchez, 2017).

For more background on the audio fingerprinting techniques mentioned here, see Haitsma & Kalker (2003).

**Box 1.**
# Deep Learning as the Foundation for Predictive Models

Deep learning is an attempt to solve complex computational problems by replicating the structure of the human brain. The result are structures called **artificial neural networks (ANNs)** that can "learn" from very large quantities of data. The basic function of ANNs is to ascertain features from inputs. For example, an ANN may learn what features of an image represent a flower, by analyzing millions of images of flowers and non-flowers. ANNs contain layers of functions (called "nodes" or "neurons") which perform various operations on the data that they are fed. The "deep" of deep learning refers to a network having many, many layers, most of which are hidden. ANNs are an *umbrella* and can contain many different types of neural networks.

Think of ANNs (very roughly) like an incredibly large imaginary car factory, larger than any currently on earth, where upwards of millions of workers process smaller components of a very complex car. Assembly of the finished project will typically be broken down into a multitude of sub-tasks. Teams of workers with specialized skills build upon the output of other workers within dedicated teams and may connect with other teams as needed. The outputs of these steps may not, by themselves, look anything like the finished product, much as an ignition coil may not be immediately recognizable as a car part (even to regular users of cars). During the process, tasks and workflow may also be shifted in real-time to make the process more efficient. Thus, someone walking through this factory would likely find it impossible to grasp the immensity of the process or the relationships between various teams and processes.

ANNs can be structured in a variety of ways. One type of ANN, a fully-connected neural network, is good at making classification decisions of simple data. This means each node in a layer is connected to all the nodes in the next layer. However, fully-connected networks suffer from computational inefficiency because they are *dense*. If the first layer contained 1,000 nodes, this would lead to 1 billion parameters after just the first layer, which will increase dramatically with dozens or hundreds of layers, or if color channels are added to the image being evaluated, for example (Elgendy, 2020). This huge number of parameters leads to high computing time, unwieldiness, and overfitting, making ANNs alone ill-suited for computer vision and audition tasks. Another type of neural network, called a convolutional neural network (CNN), seeks to address this issue.

## CONVOLUTIONAL NEURAL NETWORKS

**Convolutional neural networks (CNNs)** underlie the current most popular method for modern predictive models for content analysis. They utilize *locally connected layers* to attempt to simplify inputs to smaller representations before making classification decisions. Instead of each node being connected to every node in the previous layer and considering the entire input, nodes in a CNN consider smaller windows of the input. CNNs utilize *convolutional layers*, which act like windows (or "kernels") sliding over the input data to extract salient features by applying various *filters*. These filters perform specialized operations such as edge, contour, or corner detection for images (these operations reduce spatial dimension and resolution of the image). Then a *pooling* operation is performed where the results of the high-level features are combined. Like ANNs, CNNs have input layers, output layers, and hidden layers. Predictive models that utilize CNNs often incorporate fully-connected layers or recurrent layers for stages of their analysis.

*Here's a walkthrough of a CNN process*. Suppose a CNN is used to try to identify that an input image contains a flower. First, CNN layers will apply filters across the image to create a feature map. This means the first layers will extract very rudimentary features like edges and blobs. As these features are combined, an early layer may result in recognizing a rough outline of a flower. Another layer of features may identify a petal, stem, or leaf by their outlines, colors, and textures. Pooling layers simplify the outputs of these various feature maps. They are then "flattened" onto a long vector which expresses the data in a simplified format. These simplified outputs then can be analyzed by fully-connected layers (thus the more computationally expensive part of the calculation is now being done on a much smaller, less expensive, input) which will generate a *prediction* whether the image contains a flower. This prediction is based on the data and the model's training on images containing flowers.

▼ **Figure 10.** This graphic has been recreated, and based on an illustration of a CNN by MathWorks. Source: Learn About Convolutional Neural Networks. (2020). MathWorks. Retrieved December 17, 2020 from https://www.mathworks.com/help/deeplearning/ug/introduction-to-convolutional-neural-networks.html.

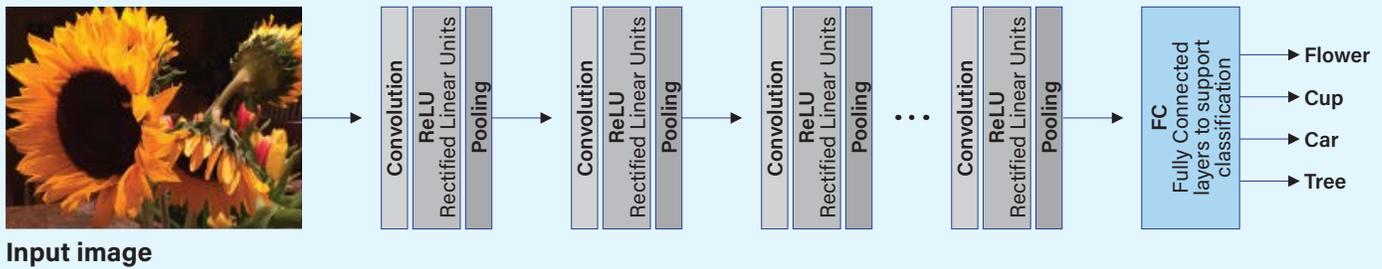

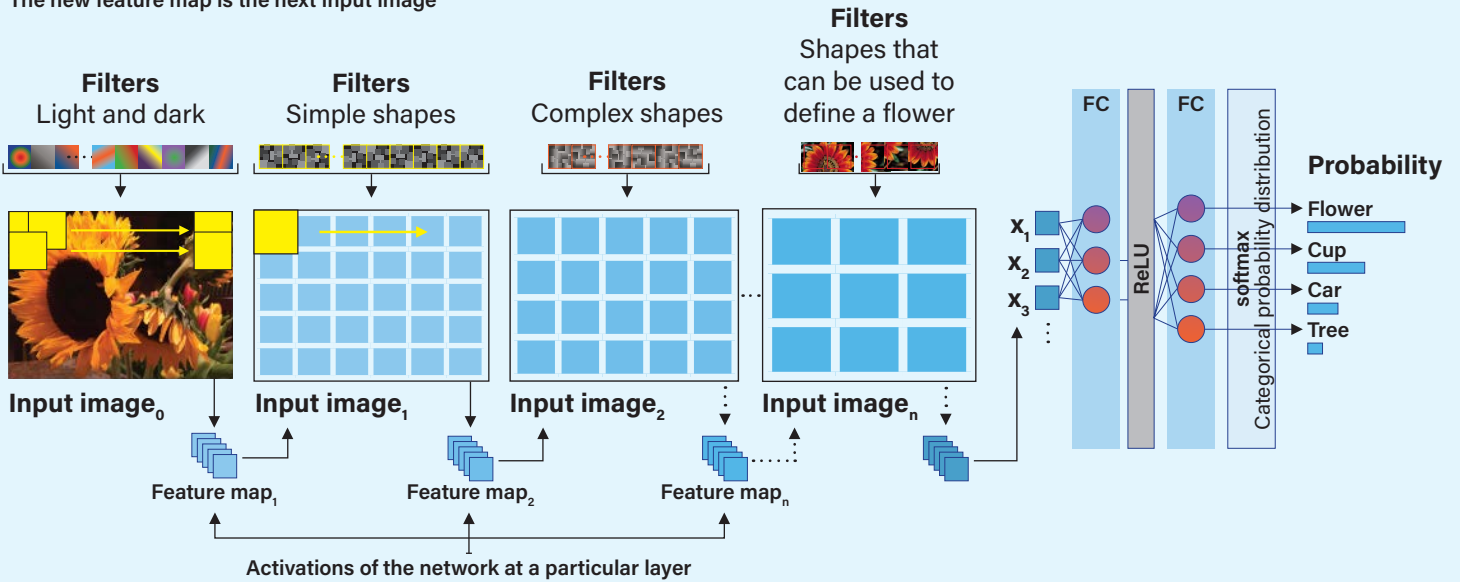

# Predictive Models - Computer Vision Models for Content Analysis

Computer vision attempts to solve a multitude of sub-problems, using techniques such as deep-learning models and CNNs. Vision problems involve a complex suite of "building block" tasks from analyzing shapes, textures, colors, spatial arrangement, and static and temporal relationships. Computer vision technology is rapidly evolving with the potential to be "the greatest disruptive innovation in a generation" (McBride, 2020). Examples of various computer vision tasks are summarized below, although this list is non-exhaustive:

| Computer Vision Task | Function | Sample Output |
| --- | --- | --- |
| **Classification** | Identifies what is in an image, without determining object location in the image. | Image contains at least one person, a hate symbol, and a sign, with a particular degree of confidence. |
| **Object Detection** | Identifies the classification and locations of objects in an image via bounding boxes. | A box-shaped region in an image contains a person, another box-region contains another person, another box contains a sign, and another box contains a hate symbol. |
| **Semantic Segmentation** | Identifies, at a pixel-level outline, what space in the image belongs to what categories of objects. | The parts of the image that are perceived as people are shaded one color, parts of the image that are signs are another color, and the hate symbol is another. |
| **Instance Segmentation** | Identifies objects using a pixel-level outline, differentiating distinct copies of the same object. | The individual people, sign, and symbol are different colors. |
| **Scene Understanding** | Identifies what is generally happening in a scene using geometric and content cues. | The scene depicts a person protesting with a sign containing a hate symbol. |
| **Action Recognition** | Identifies, using physical cues, what actions are being taken. | The person is holding the sign. The person is yelling. |
| **Object Tracking** | Identifies, in a video, where an object moves over time. | The person is swinging the sign back and forth. |
| **3D Pose Estimation** | Identifies, using joint positions, what physical action a person is taking. | The person holding the sign is making offensive gestures. |

▲ **Table 2.** Examples of various computer vision tasks summarized.

## IMAGE CLASSIFICATION (IMAGE LEVEL PREDICTIONS)

A **classifier** is a computer vision algorithm that indicates what an image contains. Image classifiers are one of the simpler computer vision tools, and they are ubiquitous and among the most common in the multimedia content analysis space (Batra, 2019). A current popular classifier is called ResNet-50, which is a CNN that contains fifty layers, is pre-trained on a million images from the ImageNet database, and can classify according to 1000 object categories.

Classification indicates what predefined categories of objects occur in data. A very basic example of a classifier would be one that *predicts* whether or not an image contains a cat or dog. "Prediction" is a term of art used since outputs are typically accompanied by a *confidence score*, which indicates the degree of certainty with which the algorithm has made its prediction (one can also think of it as a "guess").

Classifiers can achieve state-of-the-art performance across many domains. But they are brittle, meaning they are susceptible to external forms of visual interference and distortions called *perturbations* (Stock et al., 2020). Perturbations might include anything from changes to the intensity of single pixels in an image, to image-wide changes such as noise or blur. These perturbations may be environmental, a product of imperfect image capture techniques, or the result of deliberate efforts to fool an image recognition process.

▼ **Figure 11.** Illustration of image perturbations. Source: (Hendrycks & Dietterich, 2019).

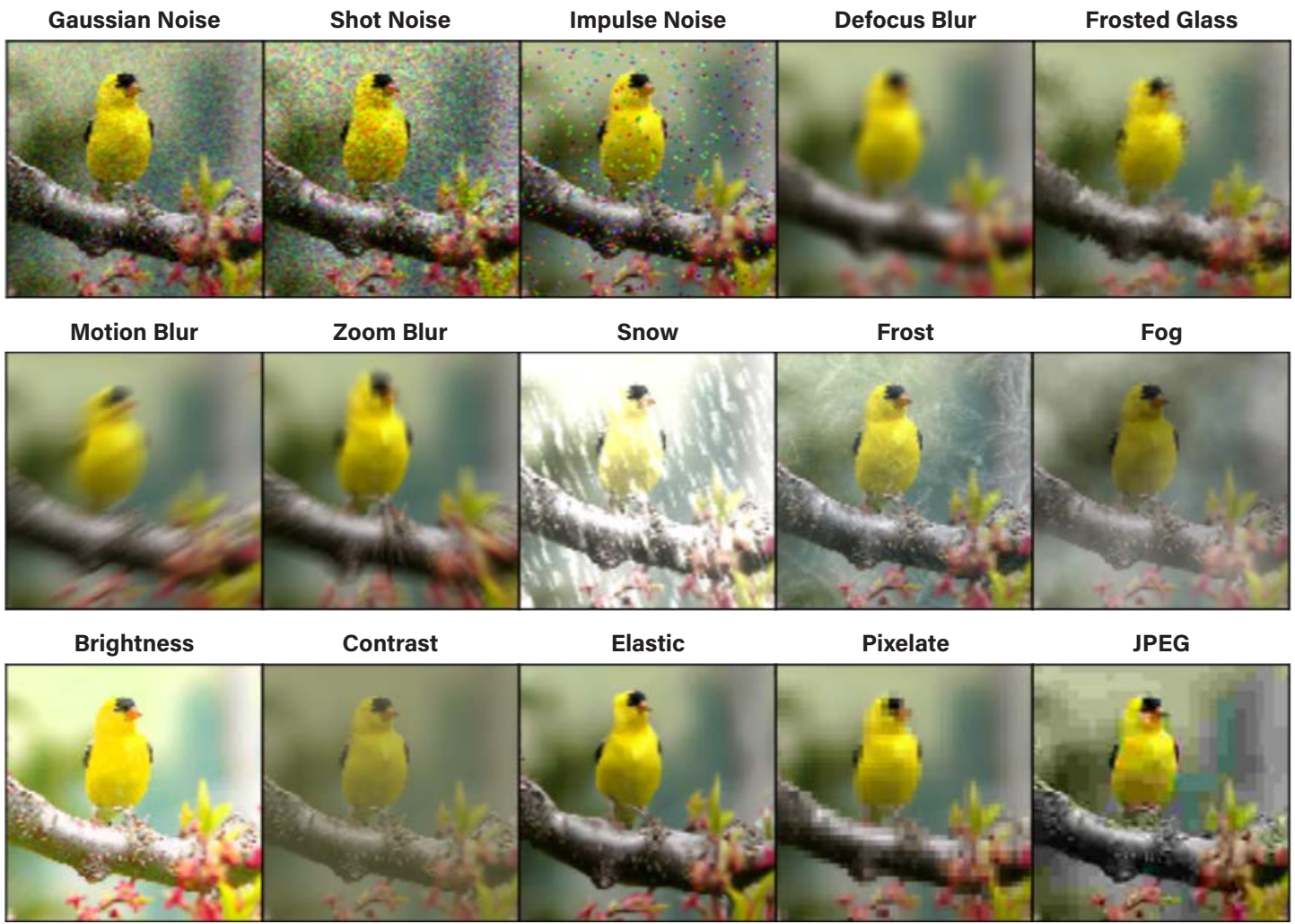

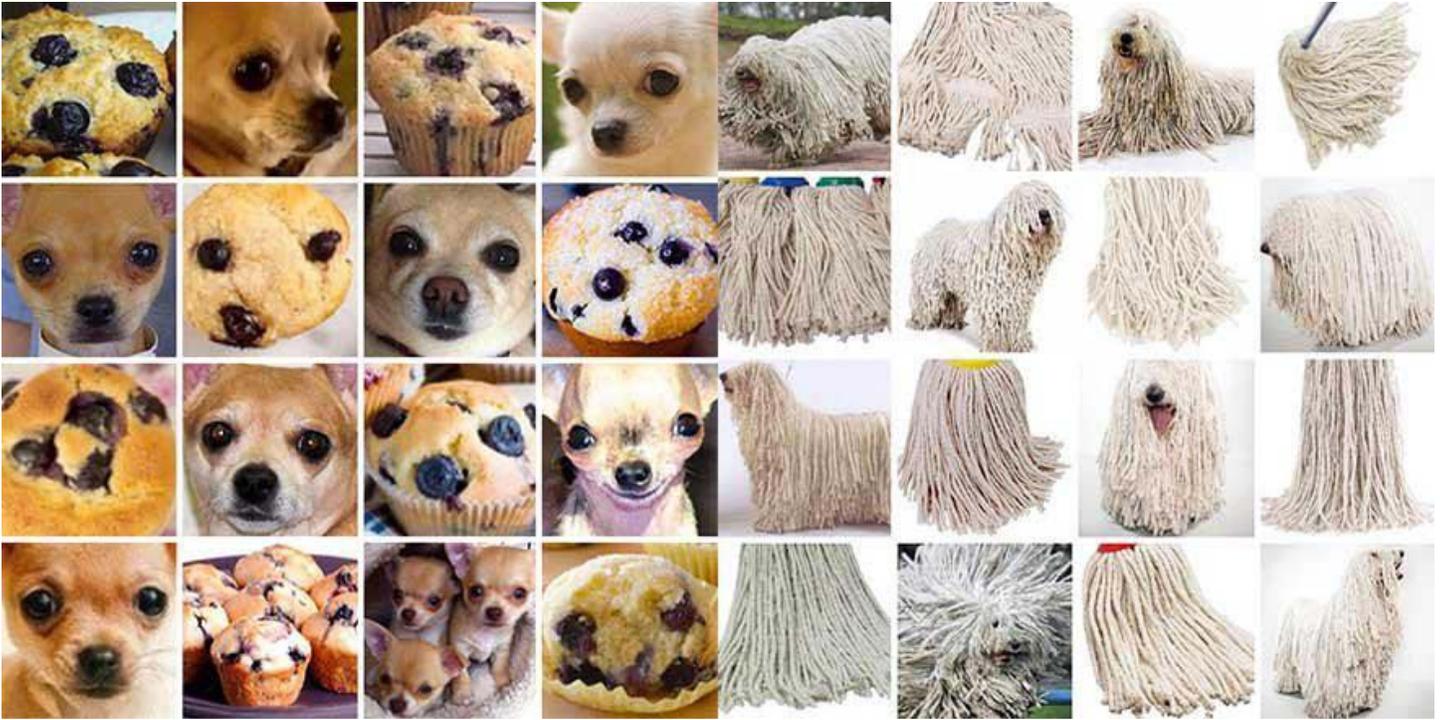

Classifiers may also be fooled by images that look very similar to one another but represent different objects, such as chihuahuas and blueberry muffins, or sheepdogs and mops (per our previous example in Figure 1).

▲ **Figure 1.** Visually similar images: chihuahuas and blueberry muffins, or sheepdogs and mops. Source: https://twitter.com/teenybiscuit/status/707670947830968320 (Accessed March 2021).

## OBJECT DETECTION (OBJECT/BOUNDING BOX LEVEL PREDICTIONS)

While classifiers merely identify what is in an image, **object detectors** take on a more complex task, which is localizing one or more objects in an image and classifying those objects. Many industry content analysis tools utilize object detectors. For instance, the Amazon Rekognition Content Moderation API, for images and videos, is a deep-learning based detector. It assigns labels to objects in photos including adult content, violence, weapons, visually disturbing content, as well as drugs, alcohol, tobacco, hate symbols, and gestures, all with associated confidence scores (Amazon Web Services, 2020). Google Cloud's Vision API similarly utilizes detectors to identify explicit content and various objects and expressions. Specialized detectors may be used in content analysis, from gunshot detectors, to blood detectors and others.

The output of a detector is typically a location, denoted by a "bounding box," and the class of the object. An object detection algorithm generally begins by proposing regions of interest (ROIs) and then conducting classification tasks, as discussed earlier, on those individual regions. Since several ROIs might initially cover an object, a process called "non-maximum suppression" is utilized to narrow down which ROI most closely frames a given object.

▶ **Figure 12.** Sample outputs of image classifiers versus detectors. This graphic has been recreated, and based on an illustration by Hulstaert, L. (2018, April 19). A Beginner's Guide to Object Detection, Datacamp. Retrieved December 17, 2020 from https://www.datacamp.com/community/tutorials/object-detection-guide.

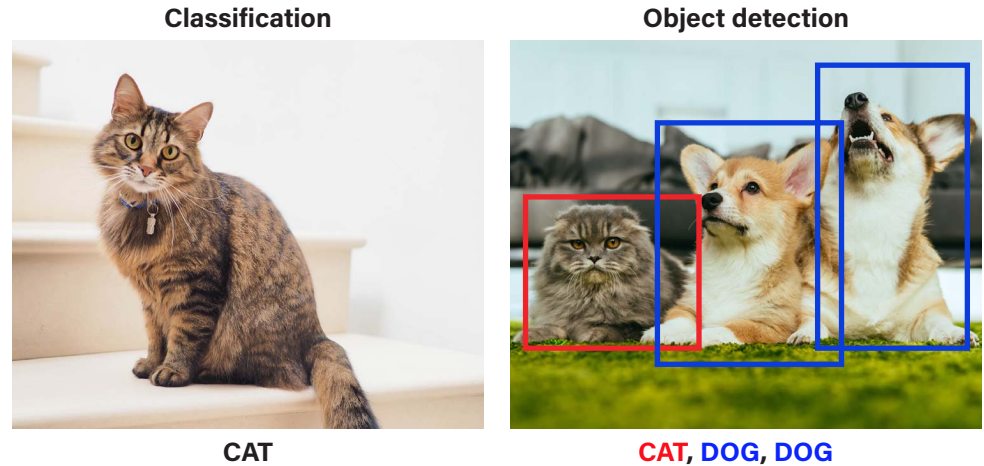

**Classification**          **Object detection**

CAT          CAT, DOG, DOG

▶ **Figure 13.** Examples of a proposal process for regions of interest (ROIs). The light green boxes would be the output ROIs because, of all the boxes, they contain the most of a given dog. This graphic has been recreated, and based on an illustration by Chanel, V.S. (2017, September 18). Selective Search for Object Detection (C++/Python), Learn OpenCV. Retrieved from https://www.learnopencv.com/selective-search-for-object-detection-cpp-python/.

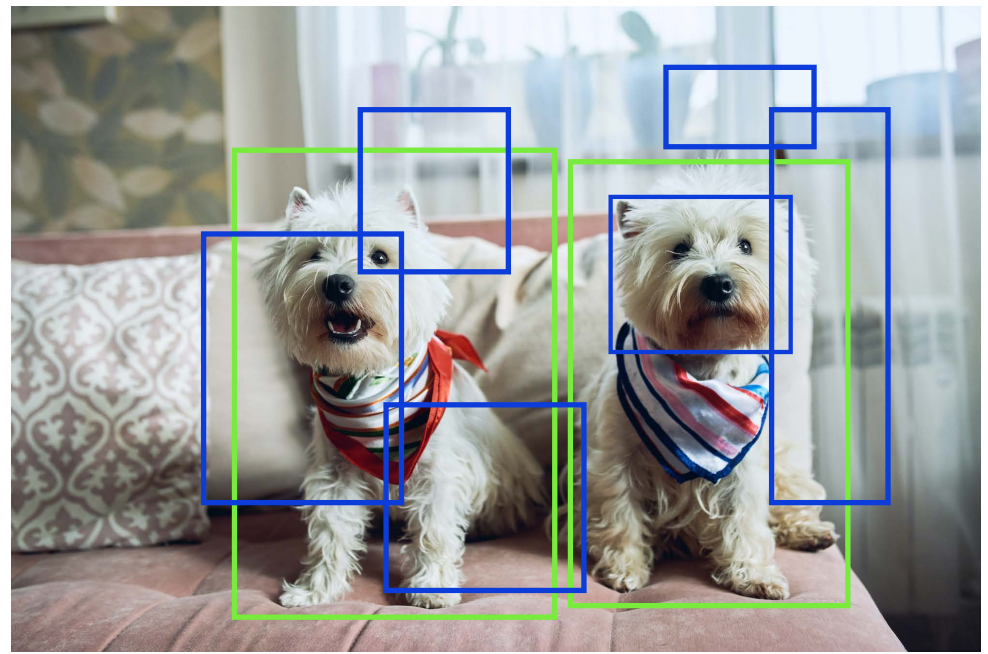

For content analysis, detectors may be desirable when the location in an image is relevant for determining its nature. For example, state-of-the-art detectors are being trained to recognize natural disasters such as earthquakes and flash floods, or other emergencies like accidents. These detectors could be used on social media to learn correlations between these events and posting metrics to better respond to emergencies (Weber et al., 2020). Object detection is crucial in analysis of video, which relies on understanding location and movement over time.

Two main evaluation metrics are used to measure the performance of object detectors. Detection speed is evaluated in frames per second (FPS), and network precision is measured via mean average precision (mAP) (Elgendy, 2020). Research shows that detectors generally perform better against efforts to circumvent them than classifiers — "fooling a detector is a very different business from fooling a classifier" (Lu et al., 2017, p. 9). This is because detectors consider a multitude of ROIs around an image, and apply a classification algorithm to each of these. Any circumvention effort must fool all of these boxes, rather than simply one. *Importantly, detectors can come in many forms, and often feature trade-offs depending on the desire for speed or accuracy.*

Improvements to R-CNN include removing the need for analysing separate region proposals (Fast R-CNN) and the use of the selective search algorithm (Faster R-CNN), both of which made computation slower (See Girshick, 2015 and ; S. Ren et al., 2016).

▼ **Figure 2.** Differences between computer vision tasks. Note that for instance segmentation, the two adjacent dogs are differentiated. In semantic segmentation, these would be the same color and not differentiated. Source: http://cs231n. stanford.edu/slides/2017/cs231n_2017_ lecture11.pdf#page=53 (Accessed May 2021).

Three of the most popular algorithms for object detection are called R-CNN, SSD (Single Shot Detector), and YOLO (You Only Look Once). R-CNN is the least sophisticated of the three. It first uses a selective search algorithm to identify the most likely regions where the object exists, runs each proposed region separately through the CNN to compute its features, and then uses a classifier to determine what the object is. These steps partly explain why the use of R-CNN architectures is slow and computationally expensive. For this reason they are called *multi-stage* detectors. SSD and YOLO attempt to address the multi-stage issue by being "one shot"—in other words, convolutional layers simultaneously predict whether ROIs contain an object while also conducting the classification step. These detectors are considerably faster, and thus are often used in real-time video or camera applications (Redmon & Farhadi, 2018). However, they tend to be more prone to mistakes than multi-stage detectors.

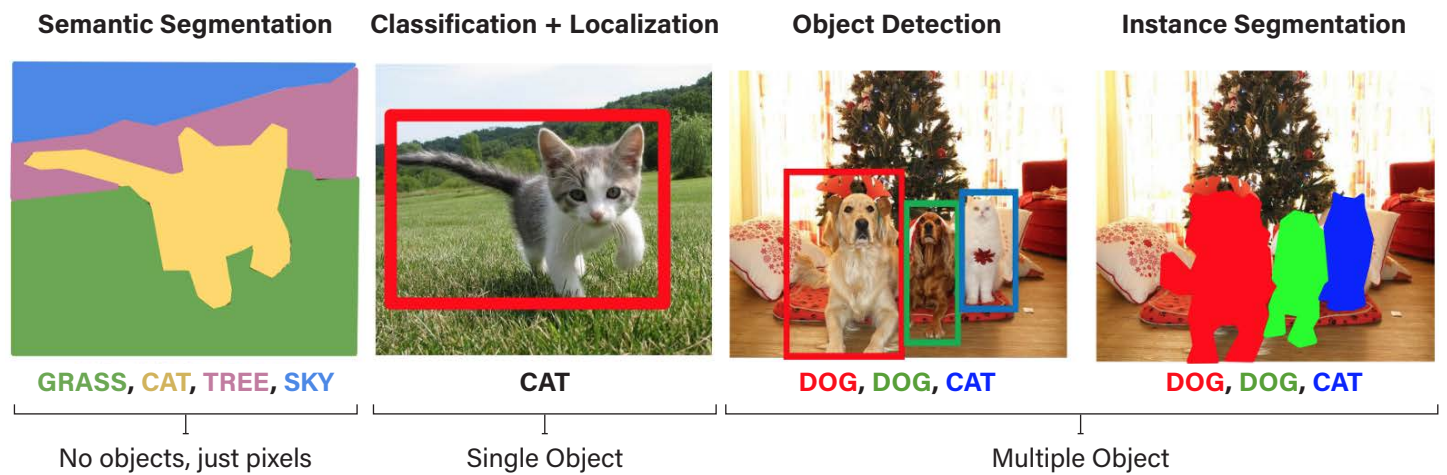

| Semantic Segmentation | Classification + Localization | Object Detection | Instance Segmentation |
| --- | --- | --- | --- |
| **GRASS, CAT, TREE, SKY** | **CAT** | **DOG, DOG, CAT** | **DOG, DOG, CAT** |
| No objects, just pixels | Single Object | Multiple Object | |

## SEMANTIC SEGMENTATION AND INSTANCE SEGMENTATION

Segmentation tasks are important for content analysis because they are the building blocks for parsing *relationships* between objects in images or video. **Semantic segmentation** seeks to be more granular than detection, by assigning a class label to each individual pixel in an image. **Instance segmentation** seeks to be even more precise and identify individual object boundaries. A popular technique for this is called Mask R-CNN, which is an extension of Faster R-CNN for object detection. It works by generating bounding boxes and then adding a step to produce "masks" or object outlines (Mittal, 2019). **Video instance segmentation** takes this further, where individually segmented instances are then linked and tracked over an entire sequence. For instance, researchers at Facebook developed an approach to instance segmentation to track objects in video sequences using a method called MaskProp. Other state-of-the-art methods in *panoptic segmentation* seek to merge both semantic and instance segmentation into one task (Kirillov et al., 2019).

## SCENE UNDERSTANDING

**Scene understanding** seeks to comprehend a scene by considering the geometric and semantic relationships of its contents (Naseer et al., 2019). Scene understanding algorithms have important applications in content analysis, as they piece together the larger correlations between individual objects. For example, an image containing "fire" might be a campfire or it could be a natural disaster or violent scene. An image containing "blood" might be a gruesome image, or it may be an educational photo of a surgery. Researchers from UCLA utilized scene understanding and visual sentiment analysis to develop a visual model to recognize protesters, describe their activities, and estimate the level of perceived violence in the image (Won et al., 2017). They identified that emotions such as anger and fear were often correlated with perceived violence, and implemented *object detection* of labels such as signs, photos, fire, law enforcement, children, and flags.

Scene understanding is a compound task that involves a number of the aforementioned "building block" tasks. Hence a scene understanding algorithm is not simply one algorithm but involves the application of a number of CNNs: classification; object detection; segmentation; monocular depth estimation; pose estimation; and / or sentiment analysis, among others.

The simplest architecture for semantic segmentation is the Fully-Convolutional Net (FCN), an encoder-decoder process. In FCN, an input image is down-sampled to a smaller size through a series of convolutions (the encoder), and then that encoded output is up-sampled. Up-sampling can occur via processes such as bilinear interpolation or transpose-convolutions (Long et al., 2015). The encoding process may, however, lead to artifacts and poor boundary resolution. More modern architectures include multi-scale models like the Pyramid Scene Parsing Network (PSPNet), which performs multiple convolution operations of varying dimensions (hence the "pyramid" title) (Zhao et al., 2017).

The MaskProp technique predicts clip-level instances in order to simultaneously classify, segment, and track object instances in video sequences. It is billed as more robust against motion blur and object occlusions in videos (Bertasius & Torresani, 2020).

## OBJECT TRACKING

The task of **object tracking** in either pre-recorded video or a live stream means following the location of a given object over time. To imagine the difficulty of this, picture being with a friend in a busy crowd. Consider the steps the brain must take to watch a friend moving through the crowd and not lose sight of them. This involves identifying individual humans in the crowd, recognizing the friend among the other humans, and differentiating the friend (or perhaps only one or more features or perspectives of the friend due to obscuration). At some moments the friend may be close or far (Asad et al., 2020). Multiple objects, lighting discrepancies, or temporary disappearances from view are just some of the problems tracking algorithms may face (Nixon & Aguado, 2019).

Video understanding is a significantly more difficult task than identification of objects in static images because it involves a temporal dimension. This dimension creates *dependencies* between various points in time (i.e., the order matters). An example of this is the act of climbing up a ladder, which can appear to be climbing down if an algorithm gets the frame-order wrong. Examples of tasks that may need to occur in video are object tracking, video object segmentation, video prediction, and pose estimation. Many current video analysis tools will approximate videos using specific frames. The Microsoft Azure content moderation system, for instance, divides content into differing "shots" and identifies specific key frames on which to run a static image analysis on whether that image is inappropriate or prohibited content.

Object tracking is utilized for a variety of use cases, such as following the motion of humans or vehicles. One key representation benefitting tracking and motion estimation is *optical flow*, or the pixel-level correspondence between images. These can help ascertain and differentiate forms of movement.

Traditionally, classical methods infer optical flow by minimizing what is called a "loss function." Modern methods utilize unsupervised learning to circumvent the need for labels. These approaches are advantageous because they yield faster results and improved performance. Examples of these approaches include OAFlow and DDFlow (Jonschkowski et al., 2020).

## ACTION RECOGNITION AND 3D POSE ESTIMATION

Advances in **action recognition** are a current priority in computer vision, given the volume of video content being produced on devices and platforms. Many action recognition algorithms are highly specialized. Tools may only consider specific subjects at a time. For example, state-of-the-art models in **violence recognition** propose to break down violence into categories such as blood, explosions, fights, fire, and firearms (Peixoto et al., 2019). **3D pose estimation** involves predicting the 3D position of human joints in images. Most reliable data is obtained using elaborate sensors and bodysuits which is impractical for collecting volumes of data and, importantly, does not exist for data obtained "in the wild" (Pavllo et al., 2019). In light of that, current research focuses on *estimation* of 3D keypoints from 2D images, historically using estimations to reference the pelvis joint. Pose estimation allows for better action recognition, as well as enabling research into human gestures. Audio cues can be combined with gestures to analyze and predict gestures from speech (Ginosar et al., 2019).

## ISSUES WITH LIVE VIDEO

Live video presents some of the most challenging problems to content analysis. It requires the application of all of the aforementioned prediction tasks. Not only must the outputs of those tasks be synthesized, but the live component requires them to be done quickly. This is enormously computationally expensive, because videos (especially high resolution ones) are large data files, and hence generally impractical to monitor for social media platforms. Use cases of screening live video for violence, for example, may thus still be far off. Facebook executives, for example, reportedly said that AI may still be years away from being able to moderate live video at scale (Kahn, 2019). However, current technologies do apply forms of live object detection. Self-driving cars must understand objects in real time (Chin et al., 2019). Even so, these technologies are typically applying detection of objects, which is a much simpler task than parsing context about whether a scene contains violence.

# Predictive Models - Computer Audition Models for Content Analysis

Computer audition seeks to understand audio content. Where audio involves humans speaking, speech will often first be *transcribed* to a text form and analyzed with natural language processing (NLP) methods. This may compound errors that are misheard (such as if "porn" is misheard as "born," potentially changing an analyzed context). Google's "AI Autobahn" combines its Natural Language API and Jigsaw's Perspective APIs to first do speech-to-text analysis, then apply textual sentiment and toxicity analysis. NLP methods and their strengths and limitations are covered in detail in CDT's previous *Mixed Messages* report (Duarte et al., 2017).

Deep learning applications for computer audition mirror many of the use cases in computer vision. However, they are typically conducted on spectrograms (graphic frequency depictions) of audio, rather than on images. State-of-the-art image classification techniques are also capable of achieving positive results on audio classification tasks (Hershey et al., 2017). Tasks of audio classification are often analogous to their image counterparts. Scene recognition, for example, has an audio counterpart (computational auditory scene recognition, or CASR) (Petetin et al., 2015). The foundational "cats and dogs" image classification task even has an audio counterpart for barks and meows (Takahashi et al., 2016).

Some unique challenges presented in computer audition include **mitigating noise**, **data variations**, and **language biases**. Isolating salient audio from noise is the subject of current research, which is attempting to isolate *sources* of audio in mixed-audio recordings (Gfeller et al., 2020). Sound samples themselves may be inconsistent, with varied loudness, sample quality, and time durations (Saska et al., 2019). Some algorithms exist for noise reduction, including spectral noise gating, which aims to eliminate consistent background noise by "gating" out any noise that falls in a certain frequency range. This can help eliminate certain types of consistent background noise, like eliminating the frequencies of a coffee grinder from a recording of ambient sounds in a coffee shop. This could be useful, for example, in a tool that is trying to identify the song playing over the coffee shop's loudspeakers. But gating out the coffee-grinder frequencies could also affect, for example, the ability of a matching algorithm to identify a song that uses those same frequencies.

Finally, **automatic speech recognition (ASR)** is challenged by the fact speech can occur in many different languages, accents, or dialects. Different recognition models may be trained on "high resource languages" (languages for which many data resources exist) versus "low resource languages" (for which there are few data resources available). Many near-extinct languages, dialects, or primarily oral languages have not generated electronic data (C. Wang et al., 2020). These considerations present challenges for the widespread application of computer audition tools for predictive applications.

# References

Agarwal, S., El-Gaaly, T., Farid, H., & Lim, S.-N. (2020). Detecting Deep-Fake Videos from Appearance and Behavior. *ArXiv:2004.14491 [Cs, Eess]*. http://arxiv.org/abs/2004.14491.

Agüera y Arcas, B., Gfeller, B., Guo, R., Kilgour, K., Kumar, S., Lyon, J., Odell, J., Ritter, M., Roblek, D., & Sharifi, M. (2017). Now Playing: Continuous low-power music recognition. *ArXiv Preprint ArXiv:1711.10958*.

Amazon Web Services. (2020, October 12). *Amazon Rekognition adds support for six new content moderation categories*. Amazon Web Services, Inc. https://aws.amazon.com/about-aws/whats-new/2020/10/amazon-rekognition-adds-support-for-six-new-content-moderation-categories/.

Amazon Web Services. (2021). *Amazon Rekognition—Developer Guide*. Amazon Web Services. https://docs.aws.amazon.com/rekognition/latest/dg/rekognition-dg.pdf#moderation.

Arya, V., Bellamy, R. K., Chen, P.-Y., Dhurandhar, A., Hind, M., Hoffman, S. C., Houde, S., Liao, Q. V., Luss, R., & Mojsilović, A. (2019). One explanation does not fit all: A toolkit and taxonomy of ai explainability techniques. *ArXiv Preprint ArXiv:1909.03012*.

Asad, H., Shrimali, V. R., & Singh, N. (2020). *The Computer Vision Workshop | Packt*. Packt. https://www.packtpub.com/product/the-computer-vision-workshop/9781800201774.

Audible Magic. (n.d.). *Patents*. Audible Magic. Retrieved April 23, 2021, from https://www.audiblemagic.com/patents/.

Barbosa, N. M., & Chen, M. (2019). Rehumanized Crowdsourcing: A Labeling Framework Addressing Bias and Ethics in Machine Learning. *Proceedings of the 2019 CHI Conference on Human Factors in Computing Systems*, 1–12. https://doi.org/10.1145/3290605.3300773.

Bartholomew, T. B. (2014). The Death of Fair Use in Cyberspace: Youtube and the Problem with Content ID. *Duke Law & Technology Review*, 13, 66.

Batra, K. (2019, June 5). *Introduction to Computer Vision and Building Applications That Can See*. https://www.youtube.com/watch?v=L2B6_s3UvZA.

Bertasius, G., & Torresani, L. (2020). Classifying, Segmenting, and Tracking Object Instances in Video with Mask Propagation. *ArXiv:1912.04573 [Cs]*. http://arxiv.org/abs/1912.04573.

Brinkman, C., Fragkiadakis, M., & Bos, X. (2016). *Online music recognition: The Echoprint system*. https://staas.home.xs4all.nl/t/swtr/documents/wt2015_echoprint.pdf.

Browning, K. (2020, November 17). Zuckerberg and Dorsey Face Harsh Questioning From Lawmakers. *The New York Times*. https://www.nytimes.com/live/2020/11/17/technology/twitter-facebook-hearings.

Buda, M., Maki, A., & Mazurowski, M. A. (2018). A systematic study of the class imbalance problem in convolutional neural networks. *Neural Networks, 106*, 249–259. https://doi.org/10.1016/j.neunet.2018.07.011.

Buolamwini, J., & Gebru, T. (2018). Gender Shades: Intersectional Accuracy Disparities in Commercial Gender Classification. *Conference on Fairness, Accountability and Transparency*, 77–91. http://proceedings.mlr.press/v81/buolamwini18a.html.

Burton-Harris, V., & Mayor, P. (2020, June 24). *Wrongfully Arrested Because Face Recognition Can't Tell Black People Apart*. American Civil Liberties Union. https://www.aclu.org/news/privacy-technology/wrongfully-arrested-because-face-recognition-cant-tell-black-people-apart/.

Cambridge Consultants. (2019). *Use of AI in online content moderation*. Cambridge Consultants. https://www.ofcom.org.uk/research-and-data/internet-and-on-demand-research/online-content-moderation.

Cavey, T., Dolan, A., & Stock, J. (2020). Strategies for Robust Image Classification. *ArXiv Preprint ArXiv:2004.03452*.

Chesney, B., & Citron, D. (2019). Deep fakes: A looming challenge for privacy, democracy, and national security. *Calif. L. Rev.*, 107, 1753.

Chin, T.-W., Ding, R., & Marculescu, D. (2019). AdaScale: Towards Real-time Video Object Detection Using Adaptive Scaling. *ArXiv:1902.02910 [Cs]*. http://arxiv.org/abs/1902.02910.

Condliffe, J. (2019, November 15). The Week in Tech: Algorithmic Bias Is Bad. Uncovering It Is Good. *The New York Times*. https://www.nytimes.com/2019/11/15/technology/algorithmic-ai-bias.html.

Counter Extremism Project. (2018). *CEP Report: YouTube's Ongoing Failure to Remove ISIS Content*. Counter Extremism Project. https://www.counterextremism.com/sites/default/files/eGLYPH_web_crawler_white_paper_July_2018.pdf.

Dalins, J., Wilson, C., & Boudry, D. (2019). PDQ & TMK+ PDQF—A Test Drive of Facebook's Perceptual Hashing Algorithms. *ArXiv Preprint ArXiv:1912.07745*.

Davis, A., & Rosen, G. (2019, August 1). *Open-Sourcing Photo- and Video-Matching Technology to Make the Internet Safer*. Facebook. https://about.fb.com/news/2019/08/open-source-photo-video-matching/.

Dolhansky, B., & Ferrer, C. C. (2020). Adversarial collision attacks on image hashing functions. *ArXiv:2011.09473 [Cs]*. http://arxiv.org/abs/2011.09473.

Douek, E. (2020, February 11). *The Rise of Content Cartels*. https://knightcolumbia.org/content/the-rise-of-content-cartels.

DrivenData. (2020). *Hateful Memes: Phase 2*. DrivenData. https://www.drivendata.org/competitions/70/hateful-memes-phase-2/page/267/.

Drmic, A., Silic, M., Delac, G., Vladimir, K., & Kurdija, A. S. (2017). Evaluating robustness of perceptual image hashing algorithms. *2017 40th International Convention on Information and Communication Technology, Electronics and Microelectronics (MIPRO)*, 995–1000. https://doi.org/10.23919/MIPRO.2017.7973569.

Du, L., Ho, A. T. S., & Cong, R. (2020). Perceptual hashing for image authentication: A survey. *Signal Processing: Image Communication*, 81, 115713. https://doi.org/10.1016/j.image.2019.115713.

Duarte, N., Llansó, E., & Loup, A. C. (2017). *Mixed Messages? The Limits of Automated Social Media Content Analysis*. Center for Democracy & Technology. https://cdt.org/wp-content/uploads/2017/11/2017-11-13-Mixed-Messages-Paper.pdf.

Eilertsen, G., Jönsson, D., Ropinski, T., Unger, J., & Ynnerman, A. (2020). Classifying the classifier: Dissecting the weight space of neural networks. *ArXiv:2002.05688v1 [Cs.CV]*. https://arxiv.org/abs/2002.05688v1.

Elgendy, M. (2020). *Deep Learning for Vision Systems*. Manning Publications.

Ellis, D., & Whitman, B. (2013). *Musical fingerprinting based on onset intervals* (United States Patent No. US8586847B2). https://patents.google.com/patent/US8586847B2/en.

Engstrom, E., & Feamster, N. (2017). *The Limits of Filtering: A Look at the Functionality & Shortcomings of Content Detection Tools*. Engine. https://www.engine.is/the-limits-of-filtering.

Eykholt, K., Evtimov, I., Fernandes, E., Li, B., Rahmati, A., Xiao, C., Prakash, A., Kohno, T., & Song, D. (2018). Robust Physical-World Attacks on Deep Learning Models. *ArXiv:1707.08945 [Cs]*. http://arxiv.org/abs/1707.08945.

Facebook. (2019, November 13). *Community Standards Report*. https://ai.facebook.com/blog/community-standards-report/.

Facebook. (2020, May 12). *Using AI to detect COVID-19 misinformation and exploitative content*. Facebook AI. https://ai.facebook.com/blog/using-ai-to-detect-covid-19-misinformation-and-exploitative-content/.

Faddoul, M. (2020, April 28). COVID-19 is triggering a massive experiment in algorithmic content moderation. *Brookings*. https://www.brookings.edu/techstream/covid-19-is-triggering-a-massive-experiment-in-algorithmic-content-moderation/.

Fawzi, A., Fawzi, H., & Fawzi, O. (2018). Adversarial vulnerability for any classifier. *ArXiv Preprint ArXiv:1802.08686*.

Geirhos, R., Rubisch, P., Michaelis, C., Bethge, M., Wichmann, F. A., & Brendel, W. (2018). ImageNet-trained CNNs are biased towards texture; increasing shape bias improves accuracy and robustness. *ArXiv Preprint ArXiv:1811.12231*. https://arxiv.org/pdf/1811.12231.

Gfeller, B., Roblek, D., & Tagliasacchi, M. (2020). One-shot conditional audio filtering of arbitrary sounds. *ArXiv:2011.02421 [Eess]*. http://arxiv.org/abs/2011.02421.

GIFCT. (n.d.). Joint Tech Innovation. *GIFCT*. Retrieved April 23, 2021, from https://gifct.org/joint-tech-innovation/#row-hash.

Ginosar, S., Bar, A., Kohavi, G., Chan, C., Owens, A., & Malik, J. (2019). Learning Individual Styles of Conversational Gesture. *ArXiv:1906.04160 [Cs, Eess]*. http://arxiv.org/abs/1906.04160.

Girshick, R. (2015). Fast R-CNN. *2015 IEEE International Conference on Computer Vision (ICCV)*, 1440–1448. https://doi.org/10.1109/ICCV.2015.169.

Goodfellow, I. J., Pouget-Abadie, J., Mirza, M., Xu, B., Warde-Farley, D., Ozair, S., Courville, A., & Bengio, Y. (2014). Generative Adversarial Networks. *ArXiv:1406.2661v1 [Stat.ML]*. https://arxiv.org/abs/1406.2661v1.

Google. (n.d.). *Fighting child sexual abuse online*. Retrieved April 23, 2021, from https://protectingchildren.google/intl/en/.

Google Cloud. (n.d.). *Vision AI*. Google Cloud. Retrieved April 23, 2021, from https://cloud.google.com/vision.

Google Cloud. (2021). *Categorizing audio content using machine learning*. Google Cloud. https://cloud.google.com/architecture/categorizing-audio-files-using-ml.

Gorwa, R., Binns, R., & Katzenbach, C. (2020). Algorithmic content moderation: Technical and political challenges in the automation of platform governance. *Big Data & Society*, 7(1). https://journals.sagepub.com/doi/full/10.1177/2053951719897945.

Greenemeier, L. (2017). When Hatred Goes Viral: Inside Social Media's Efforts to Combat Terrorism—Scientific American. *Scientific American*, 316(5). https://www.scientificamerican.com/article/when-hatred-goes-viral-inside-social-medias-efforts-to-combat-terrorism/.

Haitsma, J., & Kalker, T. (2003). A Highly Robust Audio Fingerprinting System With an Efficient Search Strategy. *Journal of New Music Research*, 32(2), 211–221. https://doi.org/10.1076/jnmr.32.2.211.16746.

Harwell, D. (2018, July 19). *The accent gap: How Amazon's and Google's smart speakers leave certain voices behind*. Washington Post. https://www.washingtonpost.com/graphics/2018/business/alexa-does-not-understand-your-accent/.

Heaven, D. (2019). Why deep-learning AIs are so easy to fool. *Nature, 574*(7777), 163–166. https://doi.org/10.1038/d41586-019-03013-5.

Hendrycks, D., Basart, S., Mu, N., Kadavath, S., Wang, F., Dorundo, E., Desai, R., Zhu, T., Parajuli, S., Guo, M., Song, D., Steinhardt, J., & Gilmer, J. (2020). The Many Faces of Robustness: A Critical Analysis of Out-of-Distribution Generalization. *ArXiv:2006.16241*. http://arxiv.org/abs/2006.16241.

Hendrycks, D., & Dietterich, T. (2019). Benchmarking neural network robustness to common corruptions and perturbations. *ArXiv Preprint ArXiv:1903.12261*. https://arxiv.org/abs/1903.12261.

Hendrycks, D., Zhao, K., Basart, S., Steinhardt, J., & Song, D. (2021). Natural Adversarial Examples. *ArXiv:1907.07174 [Cs, Stat]*. http://arxiv.org/abs/1907.07174.

Hermann, K. L., Chen, T., & Kornblith, S. (2019). The origins and prevalence of texture bias in convolutional neural networks. *ArXiv Preprint ArXiv:1911.09071*. https://arxiv.org/pdf/1911.09071.

Hershey, S., Chaudhuri, S., Ellis, D. P. W., Gemmeke, J. F., Jansen, A., Moore, R. C., Plakal, M., Platt, D., Saurous, R. A., Seybold, B., Slaney, M., Weiss, R. J., & Wilson, K. (2017). CNN Architectures for Large-Scale Audio Classification. *ArXiv:1609.09430 [Cs, Stat]*. http://arxiv.org/abs/1609.09430.

Hind, M. (2019). Explaining explainable AI. *XRDS: Crossroads, The ACM Magazine for Students*, 25(3), 16–19. https://doi.org/10.1145/3313096.

Huang, J., Rathod, V., Sun, C., Zhu, M., Korattikara, A., Fathi, A., Fischer, I., Wojna, Z., Song, Y., & Guadarrama, S. (2017). Speed/accuracy trade-offs for modern convolutional object detectors. *Proceedings of the IEEE Conference on Computer Vision and Pattern Recognition*, 7310–7311.

Human Rights Watch. (2020). *"Video Unavailable": Social Media Platforms Remove Evidence of War Crimes*. Human Rights Watch. https://www.hrw.org/report/2020/09/10/video-unavailable-social-media-platforms-remove-evidence-war-crimes.

Ilyas, A., Engstrom, L., Athalye, A., & Lin, J. (2018). Black-box Adversarial Attacks with Limited Queries and Information. *ArXiv:1804.08598*. http://arxiv.org/abs/1804.08598

Jablons, Z. (2017, May 31). Evaluating Perceptual Image Hashes at OkCupid. *OkCupid Engineering Blog*. https://tech.okcupid.com/evaluating-perceptual-image-hashes-at-okcupid-e98a3e74aa3a.

Jiang, C., & Pang, Y. (2018). Perceptual image hashing based on a deep convolution neural network for content authentication. *Journal of Electronic Imaging*, 27(4). https://doi.org/10.1117/1.JEI.27.4.043055.

Jonschkowski, R., Stone, A., Barron, J. T., Gordon, A., Konolige, K., & Angelova, A. (2020). What Matters in Unsupervised Optical Flow. *ArXiv:2006.04902 [Cs, Eess]*. http://arxiv.org/abs/2006.04902.

Kahn, J. (2019, May 17). *Facebook Executive Warns AI Video Screening Still a Long Way Off—Bloomberg*. https://www.bloomberg.com/news/articles/2019-05-17/facebook-executive-warns-ai-video-screening-still-a-long-way-off.

Karlinsky, L., Shtok, J., Alfassy, A., Lichtenstein, M., Harary, S., Schwartz, E., Doveh, S., Sattigeri, P., Feris, R., & Bronstein, A. (2020). StarNet: Towards weakly supervised few-shot detection and explainable few-shot classification. *ArXiv Preprint ArXiv:2003.06798*.

Kazakos, E., Nagrani, A., Zisserman, A., & Damen, D. (2019). EPIC-Fusion: Audio-Visual Temporal Binding for Egocentric Action Recognition. *2019 IEEE/CVF International Conference on Computer Vision (ICCV)*, 5491–5500. https://doi.org/10.1109/ICCV.2019.00559.

Kirillov, A., He, K., Girshick, R., Rother, C., & Dollár, P. (2019). Panoptic Segmentation. *ArXiv:1801.00868 [Cs]*. http://arxiv.org/abs/1801.00868.

Kozyrkov, C. (2018, May 24). *The simplest explanation of machine learning you'll ever read | Hacker Noon*. Hackernoon. https://hackernoon.com/the-simplest-explanation-of-machine-learning-youll-ever-read-bebc0700047c.

Kushwaha, A. (2019, August 24). Solving Class imbalance problem in CNN. *Medium*. https://medium.com/x8-the-ai-community/solving-class-imbalance-problem-in-cnn-9c7a5231c478.

Langston, J. (2018, September 12). How PhotoDNA for Video is being used to fight online child exploitation. *Microsoft - On the Issues*. https://news.microsoft.com/on-the-issues/2018/09/12/how-photodna-for-video-is-being-used-to-fight-online-child-exploitation/.

Lapuschkin, S., Binder, A., Müller, K.-R., & Samek, W. (2017). Understanding and Comparing Deep Neural Networks for Age and Gender Classification. *ArXiv:1708.07689v1 [Stat.ML]*. https://arxiv.org/abs/1708.07689v1.

Lee, H.-E., Ermakova, T., Ververis, V., & Fabian, B. (2020). Detecting child sexual abuse material: A comprehensive survey. *Forensic Science International: Digital Investigation*, 34, 301022. https://doi.org/10.1016/j.fsidi.2020.301022.

Llansó, E. (2016, December 6). Takedown Collaboration by Private Companies Creates Troubling Precedent. *Center for Democracy and Technology*. https://cdt.org/insights/takedown-collaboration-by-private-companies-creates-troubling-precedent/.

Llansó, E. (2020a). No amount of "AI" in content moderation will solve filtering's prior-restraint problem. *Big Data & Society*, 7(1). https://doi.org/10.1177/2053951720920686.

Llansó, E. (2020b, April 22). COVID-19 Content Moderation Research Letter—In English, Spanish, & Arabic. *Center for Democracy and Technology*. https://cdt.org/insights/covid-19-content-moderation-research-letter/.

Llansó, E. (2020c, August 21). Content Moderation Knowledge Sharing Shouldn't Be A Backdoor To Cross-Platform Censorship. *Techdirt*. https://www.techdirt.com/articles/20200820/08564545152/content-moderation-knowledge-sharing-shouldnt-be-backdoor-to-cross-platform-censorship.shtml.

Long, J., Shelhamer, E., & Darrell, T. (2015). Fully Convolutional Networks for Semantic Segmentation. *ArXiv:1411.4038 [Cs]*. http://arxiv.org/abs/1411.4038.

Lu, J., Sibai, H., Fabry, E., & Forsyth, D. (2017). Standard detectors aren't (currently) fooled by physical adversarial stop signs. *ArXiv:1710.03337 [Cs]*. http://arxiv.org/abs/1710.03337.

Lykousas, N., Gómez, V., & Patsakis, C. (2018). Adult content in Social Live Streaming Services: Characterizing deviant users and relationships. *ArXiv:1806.10577v1 [Cs.SI]*. https://arxiv.org/abs/1806.10577v1.

Lyon, J. (2018, September 14). Google's Next Generation Music Recognition. *Google AI Blog*. http://ai.googleblog.com/2018/09/googles-next-generation-music.html.

Lyons, K. (2020, September 20). Twitter is looking into why its photo preview appears to favor white faces over Black faces. *The Verge*. https://www.theverge.com/2020/9/20/21447998/twitter-photo-preview-white-black-faces.

Mack, D. (2018, April 17). This PSA About Fake News From Barack Obama Is Not What It Appears. *BuzzFeed News*. https://www.buzzfeednews.com/article/davidmack/obama-fake-news-jordan-peele-psa-video-buzzfeed.

Martínez, S., Gérard, S., & Cabot, J. (2018). Robust Hashing for Models. *Proceedings of the 21th ACM/IEEE International Conference on Model Driven Engineering Languages and Systems*, 312–322. https://doi.org/10.1145/3239372.3239405.

Matsakis, L., & Martineau, P. (2020, March 18). Coronavirus Disrupts Social Media's First Line of Defense. *Wired*. https://www.wired.com/story/coronavirus-social-media-automated-content-moderation/.

McBride, S. (2020, August 14). Introducing The Third Great Computing Superpower. *Forbes*. https://www.forbes.com/sites/stephenmcbride1/2020/08/14/introducing-the-third-great-computing-superpower/.

Microsoft. (n.d.). *Intelligible, Interpretable, and Transparent Machine Learning—Microsoft Research*. Retrieved March 21, 2021, from https://www.microsoft.com/en-us/research/project/intelligible-interpretable-and-transparent-machine-learning/.

Microsoft Azure. (n.d.). *Azure Content Moderator—Content Filtering Software*. Retrieved April 23, 2021, from https://azure.microsoft.com/en-us/services/cognitive-services/content-moderator/.

Mittal, A. (2019, June 17). Instance segmentation using Mask R-CNN. *Medium*. https://aditi-mittal.medium.com/instance-segmentation-using-mask-r-cnn-7f77bdd46abd.

Montserrat, D. M., Hao, H., Yarlagadda, S. K., Baireddy, S., Shao, R., Horváth, J., Bartusiak, E., Yang, J., Güera, D., Zhu, F., & Delp, E. J. (2020). Deepfakes Detection with Automatic Face Weighting. *ArXiv:2004.12027 [Cs, Eess]*. http://arxiv.org/abs/2004.12027.

Mu, N., & Gilmer, J. (2019). Mnist-c: A robustness benchmark for computer vision. *ArXiv Preprint ArXiv:1906.02337*. https://arxiv.org/pdf/1906.02337.

Nadeem, M. S., Franqueira, V. N. L., & Zhai, X. (2019). Privacy verification of photoDNA based on machine learning. In W. Ren, L. Wang, K. K. R. Choo, & F. Xhafa (Eds.), *Security and privacy for big data, cloud computing and applications* (pp. 263-280.). The Institution of Engineering and Technology (IET). https://derby.openrepository.com/handle/10545/624203.

Naseer, M., Khan, S. H., & Porikli, F. (2019). Indoor Scene Understanding in 2.5/3D for Autonomous Agents: A Survey. *IEEE Access*, 7, 1859–1887. https://doi.org/10.1109/ACCESS.2018.2886133.

Nie, W., Yu, Z., Mao, L., Patel, A. B., Zhu, Y., & Anandkumar, A. (2020). BONGARD-LOGO: A New Benchmark for Human-Level Concept Learning and Reasoning. *ArXiv Preprint ArXiv:2010.00763*. https://arxiv.org/pdf/2010.00763.

Nixon, M., & Aguado, A. (2019). *Feature extraction and image processing for computer vision*. Academic press.

Northcutt, C. G., Athalye, A., & Mueller, J. (2021). Pervasive Label Errors in Test Sets Destabilize Machine Learning Benchmarks. *ArXiv:2103.14749 [Cs, Stat]*. http://arxiv.org/abs/2103.14749.

Pavllo, D., Feichtenhofer, C., Grangier, D., & Auli, M. (2019). 3D human pose estimation in video with temporal convolutions and semi-supervised training. *ArXiv:1811.11742 [Cs]*. http://arxiv.org/abs/1811.11742.

Peixoto, B., Lavi, B., Martin, J. P. P., Avila, S., Dias, Z., & Rocha, A. (2019). Toward Subjective Violence Detection in Videos. *ICASSP 2019 - 2019 IEEE International Conference on Acoustics, Speech and Signal Processing (ICASSP)*, 8276–8280. https://doi.org/10.1109/ICASSP.2019.8682833.

Pereira, M., Dodhia, R., & Brown, R. (2020). Metadata-based detection of child sexual abuse material. *ArXiv Preprint ArXiv:2010.02387*. https://arxiv.org/pdf/2010.02387.pdf.

Perez, M., Kot, A. C., & Rocha, A. (2019). Detection of real-world fights in surveillance videos. *ICASSP 2019-2019 IEEE International Conference on Acoustics, Speech and Signal Processing (ICASSP)*, 2662–2666.

Petetin, Y., Laroche, C., & Mayoue, A. (2015). Deep neural networks for audio scene recognition. *2015 23rd European Signal Processing Conference (EUSIPCO)*, 125–129. https://doi.org/10.1109/EUSIPCO.2015.7362358.

Pham, L. (2019, June 12). GIFCT Webinar Reveals Little About Use of Hashing Database. *Counter Extremism Project*. https://www.counterextremism.com/blog/gifct-webinar-reveals-little-about-use-hashing-database.

Phillips, P. J., Hahn, A. C., Fontana, P. C., Broniatowski, D. A., & Przybocki, M. A. (2020). Four Principles of Explainable Artificial Intelligence (Draft). *NIST Interagency/Internal Report (NISTIR) - 8312-Draft*. https://www.nist.gov/publications/four-principles-explainable-artificial-intelligence-draft.

Powles, J. (2018, December 7). *The Seductive Diversion of 'Solving' Bias in Artificial Intelligence*. https://onezero.medium.com/the-seductive-diversion-of-solving-bias-in-artificial-intelligence-890df5e5ef53.

Prabhu, V. U., & Birhane, A. (2020). Large image datasets: A pyrrhic win for computer vision? *ArXiv:2006.16923 [Cs, Stat]*. http://arxiv.org/abs/2006.16923.

Prost, F., Qian, H., Chen, Q., Chi, E. H., Chen, J., & Beutel, A. (2019). Toward a better trade-off between performance and fairness with kernel-based distribution matching. *ArXiv:1910.11779 [Cs, Stat]*. http://arxiv.org/abs/1910.11779.

Radsch, C. (2020, September 30). GIFCT: Possibly the Most Important Acronym You've Never Heard Of. *Just Security*. https://www.justsecurity.org/72603/gifct-possibly-the-most-important-acronym-youve-never-heard-of/.

Ray, T. (2020, February 10). *AI on steroids: Much bigger neural nets to come with new hardware, say Bengio, Hinton, and LeCun*. ZDNet. https://www.zdnet.com/article/ai-on-steroids-much-bigger-neural-nets-to-come-with-new-hardware-say-bengio-hinton-lecun/.

Redmon, J., & Farhadi, A. (2018). YOLOv3: An Incremental Improvement. *ArXiv:1804.02767 [Cs]*. http://arxiv.org/abs/1804.02767.

Ren, S., He, K., Girshick, R., & Sun, J. (2016). Faster R-CNN: Towards Real-Time Object Detection with Region Proposal Networks. *ArXiv:1506.01497 [Cs]*. http://arxiv.org/abs/1506.01497.

Ringer, C., & Nicolaou, M. A. (2018). Deep Unsupervised Multi-View Detection of Video Game Stream Highlights. *ArXiv:1807.09715v1 [Cs.CV]*. https://arxiv.org/abs/1807.09715v1.

Rodehorst, M. (2019, February 1). Why Alexa won't wake up when she hears her name in Amazon's Super Bowl ad. *Amazon Science*. https://www.amazon.science/blog/why-alexa-wont-wake-up-when-she-hears-her-name-in-amazons-super-bowl-ad.

Romano, A. (2018, January 31). Why Reddit's face-swapping celebrity porn craze is a harbinger of dystopia. *Vox*. https://www.vox.com/2018/1/31/16932264/reddit-celebrity-porn-face-swapping-dystopia.

Samek, W. (2020, February 12). *Explainable AI - Methods, Applications & Recent Developments*. https://www.youtube.com/watch?v=AFC8yWzypss.

Samudzi, Z. (2019, February 11). *Bots Are Terrible at Recognizing Black Faces. Let's Keep it That Way*. https://www.thedailybeast.com/bots-are-terrible-at-recognizing-black-faces-lets-keep-it-that-way.

Sanchez, D. (2017, January 12). Will Google and YouTube Be Forced To Pull Content ID? *Digital Music News*. https://www.digitalmusicnews.com/2017/01/12/google-youtube-audible-magic-content-id/.

Sankaranarayanan, S., Balaji, Y., Jain, A., Lim, S. N., & Chellappa, R. (2018). Learning from synthetic data: Addressing domain shift for semantic segmentation. *Proceedings of the IEEE Conference on Computer Vision and Pattern Recognition*, 3752–3761.

Saska, C., DiValerio, M., & Molter, M. (2019, December 14). *Building an Audio Classifier*. https://medium.com/@anonyomous.ut.grad.student/building-an-audio-classifier-f7c4603aa989.

Scott, M., & Kayali, L. (2020, October 21). *What happened when humans stopped managing social media content*. POLITICO. https://www.politico.eu/article/facebook-content-moderation-automation/.

Shankar, S., Halpern, Y., Breck, E., Atwood, J., Wilson, J., & Sculley, D. (2017). No Classification without Representation: Assessing Geodiversity Issues in Open Data Sets for the Developing World. *ArXiv:1711.08536 [Stat]*. http://arxiv.org/abs/1711.08536.

Singh, S. (2019). *Everything in Moderation- An Analysis of How Internet Platforms Are Using Artificial Intelligence to Moderate User-Generated Content*. New America. http://newamerica.org/oti/reports/everything-moderation-analysis-how-internet-platforms-are-using-artificial-intelligence-moderate-user-generated-content/.

Snow, J. (2018, July 26). Amazon's Face Recognition Falsely Matched 28 Members of Congress With Mugshots. American Civil Liberties Union. https://www.aclu.org/blog/privacy-technology/surveillance-technologies/amazons-face-recognition-falsely-matched-28.

Solomon, L. (2015). Fair Users or Content Abusers: The Automatic Flagging of Non-Infringing Videos by Content ID on Youtube. *Hofstra Law Review*, 44, 237.

Souza, T., Lima, J. P., Teichrieb, V., Nascimento, C., da Silva, F. Q. B., Santos, A. L. M., & Pinho, H. (2018). Generating an Album with the Best Media Using Computer Vision. In A. Marcus & W. Wang (Eds.), *Design, User Experience, and Usability: Designing Interactions* (pp. 338–352). Springer International Publishing. https://doi.org/10.1007/978-3-319-91803-7_25.

Spoerri, T. (2019). On upload-filters and other competitive advantages for Big Tech Companies under Article 17 of the directive on copyright in the digital single market. *J. Intell. Prop. Info. Tech. & Elec. Com. L.*, 10, 173.

Steed, R., & Caliskan, A. (2021). Image Representations Learned With Unsupervised Pre-Training Contain Human-like Biases. *Proceedings of the 2021 ACM Conference on Fairness, Accountability, and Transparency*, 701–713. https://doi.org/10.1145/3442188.3445932.

Stock, J., Dolan, A., & Cavey, T. (2020). Strategies for Robust Image Classification. *ArXiv:2004.03452v2 [Cs.CV]*. http://arxiv.org/abs/2004.03452.

Takahashi, N., Gygli, M., Pfister, B., & Van Gool, L. (2016). Deep Convolutional Neural Networks and Data Augmentation for Acoustic Event Detection. *ArXiv:1604.07160 [Cs]*. http://arxiv.org/abs/1604.07160.

Taori, R., Dave, A., Shankar, V., Carlini, N., Recht, B., & Schmidt, L. (2020). Measuring Robustness to Natural Distribution Shifts in Image Classification. *ArXiv:2007.00644 [Cs, Stat]*. http://arxiv.org/abs/2007.00644.

Tech Against Terrorism. (2020, November 11). The Terrorist Content Analytics Platform and Transparency by Design. *VOX - Pol*. https://www.voxpol.eu/the-terrorist-content-analytics-platform-and-transparency-by-design/.

Thys, S., Van Ranst, W., & Goedemé, T. (2019). Fooling automated surveillance cameras: Adversarial patches to attack person detection. *ArXiv:1904.08653 [Cs]*. http://arxiv.org/abs/1904.08653.

Todorovic, N., & Chaudhuri, A. (2018, September 3). Using AI to help organizations detect and report child sexual abuse material online. *Google*. https://blog.google/around-the-globe/google-europe/using-ai-help-organizations-detect-and-report-child-sexual-abuse-material-online/.

Trendacosta, K. (2020). *Unfiltered: How YouTube's Content ID Discourages Fair Use and Dictates What We See Online*. Electronic Frontier Foundation. https://www.eff.org/document/unfiltered-how-youtubes-content-id-discourages-fair-use-and-dictates-what-we-see-online.

Vincent, J. (2020, October 5). *Nvidia says its AI can fix some of the biggest problems in video calls—The Verge*. https://www.theverge.com/2020/10/5/21502003/nvidia-ai-videoconferencing-maxine-platform-face-gaze-alignment-gans-compression-resolution.

Wang, C., Pino, J., & Gu, J. (2020). Improving Cross-Lingual Transfer Learning for End-to-End Speech Recognition with Speech Translation. *ArXiv:2006.05474 [Cs, Eess]*. http://arxiv.org/abs/2006.05474.

Wang, Q., Gao, J., Lin, W., & Yuan, Y. (2019). Learning from synthetic data for crowd counting in the wild. *Proceedings of the IEEE/CVF Conference on Computer Vision and Pattern Recognition*, 8198–8207. http://openaccess.thecvf.com/content_CVPR_2019/papers/Wang_Learning_From_Synthetic_Data_for_Crowd_Counting_in_the_Wild_CVPR_2019_paper.pdf.

Weber, E., Marzo, N., Papadopoulos, D. P., Biswas, A., Lapedriza, A., Ofli, F., Imran, M., & Torralba, A. (2020). Detecting natural disasters, damage, and incidents in the wild. *ArXiv:2008.09188 [Cs]*. http://arxiv.org/abs/2008.09188.

*Welcome to Echoprint*. (n.d.). Retrieved April 23, 2021, from https://echoprint.tumblr.com/how.

West, S. M., Whittaker, M., & Crawford, K. (2019). *DISCRIMINATING SYSTEMS - Gender, Race, and Power in AI* (p. 33). AI Now Institute. https://ainowinstitute.org/discriminatingsystems.pdf.

Wiggers, K. (2018, December 2). Google's Inclusive Images Competition spurs development of less biased image classification AI. *VentureBeat*. https://venturebeat.com/2018/12/02/googles-inclusive-images-competition-spurs-development-of-less-biased-image-classification-ai/.

Won, D., Steinert-Threlkeld, Z. C., & Joo, J. (2017). Protest Activity Detection and Perceived Violence Estimation from Social Media Images. *Proceedings of the 25th ACM International Conference on Multimedia*, 786–794. https://doi.org/10.1145/3123266.3123282.

Wong, Q. (2020, November 19). *Facebook's AI is flagging more hate speech before you report it*. CNET. https://www.cnet.com/news/facebooks-ai-is-flagging-more-hate-speech-before-you-report-it/.

Wu, P., Liu, J., Shi, Y., Sun, Y., Shao, F., Wu, Z., & Yang, Z. (2020). Not only Look, but also Listen: Learning Multimodal Violence Detection under Weak Supervision. *ArXiv:2007.04687v2 [Cs.CV]*. https://arxiv.org/abs/2007.04687v2.

Yuille, A. L., & Liu, C. (2019, February 9). *Limitations of Deep Learning for Vision, and How We Might Fix Them*. The Gradient. https://thegradient.pub/the-limitations-of-visual-deep-learning-and-how-we-might-fix-them/.

Zernay, R., & Hagemann, R. (2017). *ACES in the Hole? Automated Copyright Enforcement Systems and the Future of Copyright Law*. The Niskanen Center.

Zhao, H., Shi, J., Qi, X., Wang, X., & Jia, J. (2017). Pyramid Scene Parsing Network. *ArXiv:1612.01105 [Cs]*. http://arxiv.org/abs/1612.01105.

cdt.org

cdt.org/contact

Center for Democracy & Technology
1401 K Street NW, Suite 200
Washington, D.C. 20005

202-637-9800

@CenDemTech